\documentstyle[flushrt,psfig,referee]{mn}
\newif\ifAMStwofonts 
\AMStwofontstrue 
\title[Cosmological Haloes in a Low-Density Universe ]
  {The Postcollapse Equilibrium Structure of Cosmological Haloes in
a Low-Density Universe}
\author[I.T.~Iliev and P.R.~Shapiro]{Ilian T. Iliev$^{1,2}$ and  
Paul R. Shapiro$^3$\\
$^1$Dept. of Physics, The University of Texas at Austin,
          Austin, TX 78712, USA\\
$^2$Instituto de Astronom\'\i a-Universidad Nacional Autonoma de 
	M\'exico, Apdo Postal 70-264, 04510 M\'exico D.F., M\'exico,\\ 
	E-mail: iliev@astroscu.unam.mx\\
$^3$Dept. of Astronomy, The University of Texas at Austin,
          Austin, TX 78712, USA, E-mail: shapiro@astro.as.utexas.edu}
\begin{document}
\label{firstpage}
\maketitle

\begin{abstract}
An analytical model is presented for the postcollapse equilibrium
structure of virialized objects which condense out of a low-density
cosmological background universe, either matter-dominated or flat with
a cosmological constant. This generalizes the model we derived previously 
for an Einstein-de~Sitter (EdS) universe. The model is based upon the assumption
that cosmological haloes form from the collapse and virialization of
``top-hat'' density perturbations and are spherical, isotropic, and 
isothermal. This leads to the prediction of a unique, nonsingular, truncated
isothermal sphere (TIS), a particular solution of the Lane-Emden equation
(suitably modified when $\Lambda\neq0$). The size and virial temperature 
are unique functions of the mass and redshift of formation of the object
for a given background universe. The central density is roughly proportional
to the critical density of the universe at the epoch of collapse. This TIS
model is in good agreement with observations of the internal structure of 
dark matter--dominated haloes on scales ranging from dwarf galaxies to X-ray
clusters. It also reproduces many of the average properties of haloes in 
simulations of the Cold Dark Matter (CDM) model to good accuracy, suggesting that
it is a useful analytical approximation for haloes which form from realistic 
initial conditions. Our TIS model matches the density profiles of haloes in CDM 
N-body simulations outside the innermost region, while avoiding the steep 
central cusp of the latter which is in apparent conflict with observations.
The TIS model may also be relevant to nonstandard CDM models, like 
self-interacting dark matter, recently proposed to resolve this conflict.  
\end{abstract}
\begin{keywords}
cosmology: theory -- dark matter -- galaxies: clusters: general -- 
   galaxies: formation -- galaxies: haloes -- galaxies: kinematics and dynamics
\end{keywords}
\section{introduction}
Galaxies and clusters of galaxies formed when gravitational 
instability amplified
density fluctuations in the expanding cosmological background universe.
Regions denser than average eventually stopped expanding and recollapsed
to form virialized objects. 
The question of what equilibrium structure results when a density perturbation collapses 
out of the expanding background universe and virializes is central to the theory of
galaxy formation. The nonlinear outcome of the growth of Gaussian-random-noise 
cosmological density fluctuations due to gravitational instability in a hierarchical
clustering model like CDM is not amenable to direct analytical solution, however.
Instead, numerical simulations are required. As a guide to understanding these simulations, 
as a check on their accuracy, and as a means of extrapolating from simulation results 
of limited dynamic range, analytical approximations are nevertheless an
essential tool. One such tool of great utility has been the solution of the 
spherical top-hat perturbation problem (e.g. Peebles 1980, 
Padmanabhan 1993). As used in the Press-Schechter (``PS'') approximation 
(Press \& Schechter 1974) and its various refinements,
the top-hat model serves to predict well the number density of virialized haloes of 
different mass which form at different epochs in N-body simulations.
An analytical model for the internal structure (e.g. mass profile, temperature, 
velocity dispersion, radius) of these virialized haloes would be a further tool of great 
value for the semi-analytical modelling of galaxy and cluster formation, therefore.
Shapiro, Iliev \& Raga (1998; Paper I) derived such a model for an EdS universe. Here 
we shall generalize the analysis of Paper I to the case of a 
low-density universe ($\Omega_0<1$) which is either matter-dominated or flat with a
cosmological constant (i.e. $\Omega_0=1-\lambda_0$). We shall also describe how 
the TIS model can be generalized to other cosmologies with 
a nonclumping background component, such as quintessence.

It is generally assumed that the collapse to infinite density predicted by the exact,
nonlinear solution of the spherical top-hat perturbation problem is interrupted by a
violent relaxation to virial equilibrium at a finite density, as a result of the growth 
of initially small-amplitude inhomogeneities in the density distribution. 
Earlier work adopted the crude approximation that the postcollapse object which emerges
from this violent relaxation is either a uniform sphere or a singular isothermal sphere,
with the same total mass and energy as the collapsing top-hat and with a radius and
velocity dispersion (or, equivalently, gas temperature) fixed in accordance with the
virial theorem. Our first motivation, therefore, was simply to improve upon 
this earlier treatment by finding a more realistic outcome for the top-hat problem. As a 
starting point, we adopted the assumption that the final equilibrium is spherical,
isotropic, and isothermal, a reasonable first approximation to the N-body and gasdynamic
simulation results of the CDM model. The postcollapse analytical solution we
derived in Paper I from this assumption quantitatively reproduces many of the average 
properties of the haloes found in those simulations to good accuracy, so we are encouraged 
to believe that our approximation is well justified for haloes which form from
realistic initial conditions. Our model is in disagreement, however, with one detail of
the N-body simulation results that, in their very centers, simulated haloes have
cuspy profiles (e.g. Navarro, Frenk \& White 1997; ``NFW''). 
By contrast, our model predicts a density profile with
a small, uniform-density core, in good agreement
with the observed rotation curves of dark matter--dominated galaxies 
(Iliev and Shapiro 2001) and with 
cluster mass profiles inferred from gravitational lensing (Iliev and Shapiro 2000,
Shapiro and Iliev 2000). This apparent discrepancy between the cuspy profiles of the N-body   
results and the observed dark matter--dominated haloes (e.g. Moore et al. 1999) has led recently 
to a vigorous reexamination 
of the cold, collisionless nature of CDM, itself, and the suggestion that a variation
of the microphysical properties of the dark matter might make it more ``collisional'',
enabling it to relax dynamically inside these haloes so as to eliminate the central cusp
(e.g. Spergel \& Steinhardt 2000; Burkert 2000; Dav\`e et al. 2000, Firmani et al. 2000
 and references therein). 
While the details of this suggestion are still 
uncertain, our model can also be applied in that case, to the extent that we are able 
to ignore the details of the relaxation process inside the halo and approximate the
final equilibrium as isothermal.

As described in Paper I, our model is as follows:
 An initial top-hat density perturbation collapses and virializes, 
which leads to a truncated nonsingular isothermal sphere in 
hydrostatic equilibrium (TIS), a solution of the isothermal Lane-Emden 
equation. Although the mass 
and total energy of the top-hat are conserved thru collapse and virialization,
and the postcollapse temperature is set by the virial theorem (including the effect of
a finite boundary pressure), the solution is not uniquely determined by
these requirements alone. In order to find a unique solution, some additional information 
is required. We adopted the anzatz that the solution selected by nature is 
the ``minimum-energy solution'' such that the boundary pressure is that for which the 
conserved top-hat energy is the minimum possible for an isothermal sphere 
of fixed mass within a finite truncation radius. This assumption fixes the 
ratio of the radius of the postcollapse sphere to the radius of the
top-hat perturbation which created it, as measured at the latter's epoch of 
maximum expansion, uniquely. For comparison, we appealed to
the details of the exact, self-similar, spherical, cosmological infall 
solution of Bertschinger (1985). In this solution, an initial overdensity
causes a continuous sequence of spherical shells of cold matter, both 
pressure-free dark matter and baryonic fluid, centered on the overdensity,
to slow their expansion, turn around and recollapse. The baryonic infall
is halted by a strong accretion shock, while density caustics form in the
collisionless dark matter, instead, due to shell-crossing. The postcollapse, 
virialized object we wish to model is then identified with the spherical 
region of shell-crossing dark matter and shock-bounded baryonic fluid in 
this infall solution (for ratio of specific heats $\gamma=5/3$) for which 
the mass is the same as that of our top-hat and the trajectory of its 
outermost spherical mass shell was identical to that of the outer boundary
of our collapsing top-hat at all times until it encountered the shock.
According to the detailed similarity solution for this infall problem, 
this spherical region of post-shock gas and shell-crossing dark matter,
it turns out, 
is very close to hydrostatic and isothermal and has virtually the same 
radius as that of the minimum-energy solution for the matching TIS. This 
offers some support for
our ``minimum-energy'' anzatz and explains the dynamical origin of the 
boundary pressure as that which results from 
thermalizing the kinetic energy of infall. 

With this ``minimum-energy'' anzatz, we found that top-hat perturbation 
collapse leads to a unique, nonsingular TIS, which yields a universal, 
self-similar density profile for the  postcollapse equilibrium of virialized
cosmological haloes. Our solution  has a unique length scale and amplitude set by the top-hat 
mass and collapse epoch, with a density at every point which is
proportional to the background density
at that epoch. The density profiles for gas and dark matter are
assumed to be the same (no bias). The final virialized halo has a small but
flat-density core.
For the EdS case, this core radius $r_0$ is about 1/30 of the total size $r_t$ of the
halo (i.e. the truncation radius). [We note that this core radius 
$r_0\equiv r_{0\rm,King}/3$, where $r_{0\rm,King}$ is the ``King radius'' defined 
by Binney \& Tremaine (1987, equ. [4-124b])].
The central density $\rho_0$ is about 500 times the density $\rho_t$ at the surface 
and 18,000 times the background density at the collapse epoch. At intermediate radii, 
$\rho$ drops faster than $r^{-2}$, as fast as $r^{-2.5}$.
The one-dimensional velocity dispersion $\sigma_V$ of the dark matter and the gas 
temperature $T$ are given simply by $\sigma_V^2=k_BT/m=4\pi G\rho_0r_0^2$,
where $m$ is the mean gas mass per gas particle.
This temperature is significantly different from that predicted by 
the standard uniform sphere (SUS) and singular isothermal sphere (SIS) 
approximations adopted previously in the literature, with
$T=2.16\,\, T_{\rm SUS}=0.72\,\, T_{\rm SIS}$.

The derived mass profile of our TIS solution as a function of its virial temperature 
(or velocity dispersion) and its formation epoch reproduces to remarkably 
high accuracy (i.e. of order 1\%) 
the cluster mass-radius-temperature relationships and average mass profile
for CDM haloes
which Evrard, Metzler, and Navarro (1996) derived
empirically by fitting a large set of detailed numerical gas 
dynamical and N-body simulation results of cluster formation in the CDM
model. The TIS and NFW halo mass profiles are also in very close agreement 
(fractional deviation of less than $\sim10\%$) 
at all radii outside of a few TIS core radii (i.e. about one King radius) 
for NFW concentration parameters $4\leq c_{\rm NFW}\leq7$.
In short, our TIS model is in good agreement with the average
properties of simulated CDM haloes 
although it differs from these numerical results
at very small radii.

The purpose of this paper is to generalize the results of Paper I to 
the case of a low-density universe, either open and matter-dominated 
($\Omega_0<1$, $\lambda_0=0$), or flat with cosmological constant 
($\Omega_0=1-\lambda_0$). In the former case, the internal structure 
of the TIS haloes will be the same as for the EdS results of Paper I,
when radius and density are expressed in dimensionless terms in
units of the core radius and central density. However, these latter 
dimensional quantities will be functions of the total mass and 
the formation epoch of the halo, functions which differ from the EdS 
results and depend upon the value of $\Omega_0$. These differences 
between the EdS solution for the TIS model in Paper I and that 
presented here for the open, matter-dominated universe ($\Omega_0<1$) 
can be understood as a reflection of the different solutions for the 
spherical top-hat problem in the two cases. For the case with a 
cosmological constant, however, the differences with respect to the 
EdS solution are much more extensive. In this case, not only is it 
necessary to consider the differences which result from the different 
solution for the spherical top-hat problem, but it is also necessary to 
take proper account of nonzero $\Lambda$ in the virial theorem, the 
conservation of energy, and the isothermal Lane-Emden equation, as well. In 
\S~\ref{top_low_dens}, we shall summarize the spherical top-hat perturbation 
problem generalized to these low-density Friedmann universe cases. We
will also briefly describe how the conservation of energy and the virial theorem
are generalized in the presence of some uniform background component X  
of energy density, such as the cosmological constant or quintessence, and
how these are combined with the top-hat solution to derive the properties 
of the postcollapse virialized object when the latter is assumed to be a 
uniform sphere [i.e. the ``standard uniform sphere approximation'' 
in the presence of the X-component]. We will then 
specialize these general results to
the case in which the uniform component X is the cosmological constant.
In \S~\ref{isoth_X}, we will generalize the Lane-Emden equation for an 
isothermal sphere in hydrostatic equilibrium to take account of the 
presence of the X-component and specialize our discussion of its solutions 
to the cosmological constant case. In \S~\ref{virial_low}, we shall apply the
virial theorem to these generalized isothermal Lane-Emden spheres,
taking care to account properly for the important effect of finite 
boundary pressure. In \S~\ref{min_en}, we will derive the 
minimum-energy TIS solutions for these low-density universe cases, including 
convenient analytical fitting formulae for practical application, in which 
all properties of the virialized post-collapse object are given as 
explicit functions of the mass and collapse redshift of the object for
a given background cosmology. Our results
and conclusions are summarized in \S~\ref{summary},
where we compare the TIS solutions for low-density universes 
derived here with the SUS and SIS approximations for those cases, as well
as with the TIS solution of Paper I for the EdS case. Finally, in 
view of the suggestive
physical correspondence between our TIS solution in the EdS case and 
the shock- and caustic-bounded sphere in the self-similar, spherical infall
solution of Bertschinger (1985), we have added an Appendix~\ref{appendix1} in which we 
show how the latter solution (for an EdS universe) can also be 
applied at early times in the generalized low-density universe cases 
described here, by a simple rescaling of parameters. 
 

\section{Spherical Top-Hat Perturbations in a Low-Density Friedmann Universe
and the Standard Uniform Sphere Approximation}
\label{top_low_dens}

\subsection{Before Collapse: the Exact  Nonlinear Solution}
\label{before}
The spherical top-hat model, an uncompensated spherical perturbation of 
uniform overdensity within a finite radius (Gunn \& Gott 1972), 
 affords considerable insight into the 
dynamics of the gravitational growth of cosmic structure, while still having an 
exact, analytical or semi-analytical solution (e.g. Peebles 1980). 
In what follows, we shall consider top-hat perturbations
in cosmological models with two components: (1)
a nonrelativistic component, which comprises all forms of matter,
luminous or dark, baryonic or non-baryonic, that can cluster under 
the action of gravity, and (2) a uniform component, which does not clump
at any scale of interest. As a special case we shall consider low-density
matter-dominated universes with no uniform component, as well.

The equation of state of the uniform component X relates its pressure
$p_X$ to its energy-density $\rho_X$ according to
\begin{equation}
\label{state}
\displaystyle{p_X=\left(\frac n3-1\right)\rho_X c^2\equiv w\rho_X c^2}.
\end{equation}
Let us also define the quantity 
\begin{equation}
\label{rho_eff}
\rho_{\rm X,eff}\equiv \rho_X+3p_X/c^2=(1+3w)\rho_X=(n-2)\rho_X.
\end{equation}
The mean rest mass energy density $\rho_b$ of the nonrelativistic component 
and the energy density $\rho_X$ then each vary with time according to
\begin{eqnarray}
\label{rho_bg_low}
\rho_b(t)&\propto&a(t)^{-3}\,,\\
\label{rho_X_low}
\rho_X(t)&\propto&a(t)^{-n}\,,
\end{eqnarray}
where $a(t)$ is the Robertson-Walker scale factor, and
$n$ is a non-negative constant. [For details and references, see e.g.
 Martel \& Shapiro (1998), and Wang \& Steinhardt (1998)].
Particular values of~$n$ correspond to models 
with a nonzero cosmological constant ($n=0$, $w=-1$), domain walls 
($n=1$, $w=-2/3$), string networks ($n=2$, $w=-1/3$), vacuum stress 
or massive neutrinos ($n=3$, $w=0$), radiation background ($n=4$,
$w=4/3$), and quintessence ($0\leq n<3$, $-1\leq w<0$). 

The time-evolution of the scale factor~$a(t)$ is described by the 
Friedmann equation. For the two-component models considered here, this 
equation takes the form
\begin{equation}
\begin{array}{ll}
\label{Friedmann}
\displaystyle{\biggl({\dot a\over a}\biggr)^2\equiv H(t)^2}
=\displaystyle{H_0^2\Biggl[
(1-\Omega_0-\Omega_{\rm X,0})\biggl({a\over a_0}\biggr)^{-2}\Biggr.}
\displaystyle{+\Omega_0\biggl({a\over a_0}\biggr)^{-3}+
\Omega_{\rm X,0}\biggl({a\over a_0}\biggr)^{-n}\Biggr]},
\end{array}
\end{equation}
where $H(t)$ is the Hubble parameter, the density parameters of the 
nonrelativistic matter
and X components are $\Omega=\rho_b/\rho_{\rm crit}$ and 
$\Omega_X=\rho_X/\rho_{\rm crit}$, respectively, where 
$\rho_{\rm crit}=3H^2(t)/8\pi G$, and all subscripts ``0'' refer to present values.
The redshift $z$ that corresponds to a scale factor $a(t)$ is given by the
usual relation
\begin{equation}
\label{z-a}
1+z=\frac{a_0}{a(t)}.
\end{equation}
In what follows, it will be convenient to define the scale factor at present by
\begin{equation}
\label{a0}
a_0\equiv\cases{
\displaystyle{\left(\frac{\Omega_{\rm X,0}}{\Omega_0}\right)^{1/(3-n)}}
,&$\rho_X\neq0\,;$\cr
1\,,&matter-dominated, $\Omega_0=1\,;$\cr
\displaystyle{\frac{1-\Omega_0}{\Omega_0}}
		\,,&matter-dominated, $\Omega_0<1\,$.\cr}
\end{equation}

While the X-component does not clump under the force of gravity, its
presence not only affects the the rate of expansion of the universe
as indicated by equation~(\ref{Friedmann}); it also modifies the 
gravitational forces on the matter component, as follows.
The Poisson equation in the presence of this non-clumping component is 
\begin{equation}
\label{Poisson}
{\bf \nabla}^2\Phi_{\rm tot}=4\pi G(\rho+\rho_{\rm X,eff}),
\end{equation}
and the corresponding gravitational potential is
\begin{equation}
\label{potential}
\Phi_{\rm tot} = \Phi + \frac{2\pi G{\rho}_{\rm X,eff}r^2}3
\equiv\Phi+\Phi_{\rm X}, 
\end{equation}
where $\Phi$ is the gravitational potential due to the nonrelativistic matter 
component as it would be in the absence of
 the X-component, and $\Phi_{\rm X}$ is the correction due to the X-component.
From Birkhoff's theorem, the collapse of a spherical top-hat 
density perturbation can be described by the Friedmann equation
for a universe with a higher average density than that of the background universe 
outside the top-hat. Special cases of the top-hat 
model in universes with a nonclumping component were discussed in  Lahav et al.
 \shortcite{LLPR} (for the cosmological constant), and 
Wang \& Steinhardt (1998) (for quintessence).
Here we shall only give the basic equations and refer the reader to 
these papers for further details and references.

The density inside the top-hat stays uniform during collapse. We can, 
therefore, describe its evolution solely in terms of its overdensity $\delta$ with 
respect to the background. 
The equation for the evolution of the top-hat overdensity in the presence of
a nonclumping component with $n<3$ is (Shapiro, Martel, and Iliev 2001)
\begin{eqnarray}
\label{delta_gen_low}
\displaystyle{\frac{d^2\,\ln(\delta +1)}{d\ln a^2}
      -\frac13\left(\frac{d\,\ln(\delta +1)}{d\ln a}\right)^2}
   +\displaystyle{\frac{g_1}{g_2}\frac{d\,\ln(\delta +1)}{d\ln a}}
		-\displaystyle{\frac{3(\delta +1)}{g_2}}=0,
\end{eqnarray}
 where 
\begin{eqnarray}
\label{g1_g2}
g_1\equiv 2\kappa a+1+(4-n)a^{3-n},\,\,\,
g_2\equiv 2(\kappa a+1+a^{3-n}),\,\,\,
\kappa\equiv \frac{(1-\Omega_0-\Omega_{X,0})\Omega_0^{(n-2)/(3-n)}}
		{\Omega_{X,0}^{1/(3-n)}}.
\end{eqnarray}
The boundary conditions are given by
\begin{equation}
\displaystyle{\delta(0)=0,\qquad \frac{d\delta}{da}(0)=A,}
\end{equation}
where $A$ indicates the initial amplitude of the top-hat fluctuation.
For $\kappa$=0, the solution $\delta(a)$ of equation (\ref{delta_gen_low})
is independent of $\Omega_0$ and $\Omega_{\rm X,0}$. In that case, we
need solve equation~(\ref{delta_gen_low}) only once to obtain the family
of solutions for different $\Omega_0$.
For open, matter-dominated models,
equation~(\ref{delta_gen_low}) is replaced by
\begin{eqnarray}
\label{delta_open}
\displaystyle{\frac{d^2\,\ln(\delta +1)}{d\ln a^2}}
-\displaystyle{\frac13\left(\frac{d\,\ln(\delta +1)}{d\ln a}\right)^2}
	+\displaystyle{\frac{2a+1}{2(a+1)}\frac{d\,\ln(\delta +1)}{d\ln a}
		-\frac{3\delta}{2(a+1)}}=0
\end{eqnarray}
(Shapiro et al. 2001).

Let us define the critical density contrast $\delta_{\rm crit}$ as the solution 
$\delta_L$ of the linearized version of equation (\ref{delta_gen_low}) or
equation (\ref{delta_open}),
extrapolated to  the epoch at which the nonlinear solution $\delta$ predicts an 
infinite overdensity. This quantity $\delta_{\rm crit}$ labels the time of 
collapse in scale-free units. The values obtained at turnaround 
will be denoted by subscript ``$m$'' for ``maximum expansion,''
while the values at collapse time will be denoted by subscripts ``coll''. 
Henceforth, we shall refer to $z_{\rm coll}$ as the redshift which corresponds
to the epoch of infinite collapse, at which $\delta=\infty$, at time 
$t_{\rm coll}$.

\subsection{After Collapse: The Standard Uniform Sphere 
Approximation}
\label{SUS}

The properties of the final, virialized object which is postulated to result from
top-hat collapse were derived in the ``standard uniform sphere 
approximation'' 
for the case with a cosmological constant 
by Lahav { et al.} \shortcite{LLPR} and for the case of quintessence
by Wang \& Steinhardt \shortcite{WS}. The postcollapse state is fully 
described in this approximation by
the radius and internal velocity dispersion of the final equilibrium sphere.  

The SUS approximation assumes that the 
collapse of the top-hat to infinite density is interrupted by a 
rapid equilibration when $\delta_L=\delta_{\rm crit}$,
which results in another uniform sphere in virial equilibrium.
The final radius $r_{\rm vir}$ of the virialized sphere is obtained
by assuming that the energy of the top-hat is conserved during
collapse and virialization and applying the virial theorem to the 
final state, assuming that the boundary pressure term in the virial theorem
can be neglected. The conserved total energy of the sphere is $E=K+W_{\rm tot}$,
where $K= U_{\rm th}+T_{\rm kin}$, $U_{\rm th}$ and $T_{\rm kin}$ 
are the thermal and kinetic energy, respectively,  
\begin{eqnarray}
\label{poten_en}
W_{\rm tot}&=& \frac 12\int_V\rho{\bf\nabla}\Phi_{\rm tot}\cdot{\bf r} dV
		\equiv W + W_{\rm X}
\end{eqnarray}
is the total potential energy in the presence of the X-component, the terms 
$W$ and $W_X$ are defined by the integral expression in equation (\ref{poten_en})
with $\Phi$ and $\Phi_X$, respectively, and 
$\Phi_{\rm tot}$, $\Phi$ and $\Phi_X$ are defined by equation~(\ref{potential}).

Let us define the collapse factor which relates the size of the sphere at
maximum expansion $r_m$ to its postcollapse equilibrium size $r_{\rm vir}$ as
$\eta\equiv r_{\rm vir}/r_m$. The relative gravitational importance of the
X-component and the matter component at maximum expansion are indicated 
by the dimensionless ratio
$\theta\equiv-(\rho_{\rm X,eff}/2\rho)_m$, where $\rho$ is the matter density 
inside the top-hat.
At the point of maximum expansion [which, unlike the EdS case, 
exists only for top-hat perturbations of amplitude high enough to enable 
collapse (see, for example, Martel 1994)], the
sphere is cold and at rest, so $K=0$, and its energy consists entirely of
gravitational potential energy.
For a uniform sphere of mass $M_0$ and radius $r_m$,  
equations~(\ref{potential}) and (\ref{poten_en}) yield
\begin{eqnarray}
\label{en1_low}
E = (W+ W_X)_m= - \frac 35 \displaystyle{\frac {GM_0^2}{r_m}}
	\left(1-\frac{\rho_{\rm X,eff}}{2\rho}\right)_m
	=  - \frac 35 \displaystyle{\frac {GM_0^2}{r_m}}(1+\theta).
\end{eqnarray}
After collapse, when the system settles down to a virial equilibrium, 
the total potential energy is 
\begin{eqnarray}
\label{old_en}
(W+W_X)_{\rm vir} 
	= - \frac 35 \displaystyle{\frac {GM_0^2}{r_{\rm vir}}}
	\left(1-\frac{\rho_{\rm X,eff}}{2\rho}\right)_{\rm vir}
	= - \frac 35 \displaystyle{\frac {GM_0^2}{r_{\rm vir}}
 \left[1+\theta\eta^3\displaystyle{\left(\frac{a_{\rm coll}}{a_m}\right)^{-n}}\right]},
\end{eqnarray}
where $a_m$ is the scale factor at turnaround and $a_{\rm coll}$ is the 
scale factor at collapse.
According to the virial theorem, the kinetic and gravitational potential
energies are related at this epoch by 
\begin{equation}
3(\gamma-1)K+W-2W_{\rm X}=0,
\end{equation}
where $\gamma$ is the ratio of specific heats. We take $\gamma=5/3$.
Together with the conservation of energy, this implies that
$E=W/2+2W_{\rm X}$,
and, therefore, the total energy in equation~(\ref{en1_low}) is related 
to the virial radius and densities after collapse according to
\begin{equation}
\label{en2_low}
E = - \frac 3{10} \displaystyle{\frac {GM_0^2}{r_{\rm vir}}
	\left(1-\frac{2\rho_{\rm X,eff}}{\rho}\right)_{\rm vir}}.
\end{equation}

Equating the total energy $E$ in equations 
(\ref{en1_low}) and (\ref{en2_low}) and using equation~(\ref{rho_X_low}) 
then yields an equation for the collapse factor $\eta$,
\begin{equation}
\label{eta_SUS}
4\theta\eta^3\displaystyle{\left(\frac{a_{\rm coll}}{a_m}\right)^{-n}}
		-2(1+\theta)\eta+1 = 0.
\end{equation}

Hereafter we shall limit our discussion to the case $n=0$ (cosmological constant),
with open, matter-dominated models as a limiting case for which 
$\rho_{\rm X,eff}=0$. Equation (\ref{eta_SUS}) then becomes
\begin{equation}
\label{eta_SUS0}
4\theta\eta^3-2(1+\theta)\eta+1 = 0,
\end{equation}
where $\theta=\rho_\lambda/\rho_m$ and $\rho_\lambda$ is the constant vacuum 
energy density associated with the cosmological constant [i.e. 
$\rho_\lambda=\Lambda/(8\pi G)$].
For any matter-dominated universe, $\theta =0$ and equation (\ref{eta_SUS0}) reduces to
the well-known relation $\eta=1/2$.
In general, equation (\ref{eta_SUS0}) has a closed form solution $\eta(\theta)$. However
the expression is quite complicated and not very practical. Instead, simpler,
approximate solutions have been proposed (Lahav et al. 1991,
Kochanek 1995). We have derived our own, better approximation to the exact
 solution [with an error of order $O(\theta^5)$]:
\begin{equation}
\label{eta_SUS_approx}
\eta = 0.5-0.25\theta-0.125\theta^2+0.125\theta^3+0.21875\theta^4.
\end{equation}

\begin{figure}
\centerline{\psfig{figure=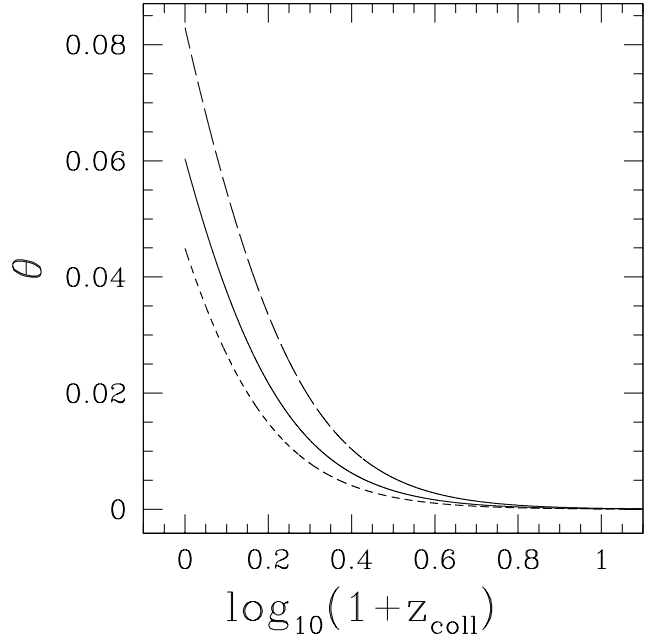,height=3.5in,width=3.5in}}
\caption{Dimensionless parameter $\theta=\rho_\lambda/\rho_m$, the ratio of the
cosmological constant energy density to matter rest-mass energy density inside
the top-hat at maximum expansion, versus the collapse redshift of
 the top-hat, $z_{\rm coll}$, for different flat background universes, as labeled
with the values of $\lambda_0$: 
$\lambda_0=0.6$ (short-dashed line), $\lambda_0=0.7$ (solid curve), and
$\lambda_0=0.8$ (long-dashed line).}
\label{theta_zcoll}
\end{figure}

The upper limit for the $\theta$ parameter is $1/2$.
If $\theta>1/2$, then the right hand side of equation~(\ref{Poisson}) 
evaluated at the epoch of maximum expansion would be negative, and,
therefore, the
perturbation would not be bound. For $\theta\rightarrow1/2$ from below, the
top-hat perturbation is only marginally bound and would collapse 
arbitrarily far into the future. 
Plots of $\theta$ vs. $z_{\rm coll}$ for several different values of 
$\lambda_0=1-\Omega_0$ are shown in Fig.~\ref{theta_zcoll}.

The virial temperature and velocity dispersion in this SUS approximation
are derived as follows. The kinetic energy in the virialized state is 
\begin{equation}
\label{K}
K=E-W_{\rm tot} 
= \frac 3{5} \displaystyle{\frac {GM_0^2}{r_{\rm m}}[1+\theta(1-3\eta^2)]}.
\end{equation}
This kinetic energy is the energy of random internal motions only 
($T_{\rm kin}=0$).
To obtain the equivalent temperature, we express the thermal energy as
\begin{equation}
\label{Uth}
U_{\rm th}=\frac 32\displaystyle{ \frac{k_B T}{m}M_0},
\end{equation}
where $m$ is the mean mass per gas particle. 
If $m_H$ is the mass of a hydrogen 
atom, then $m=\mu m_H$, where $\mu$ is the mean molecular weight.
The virial temperature is, therefore, given by equations (\ref{K}) and
(\ref{Uth}) as
\begin{equation}
\label{temp_low}
T =\displaystyle{\frac 2 5\frac {GM_0m}{k_Br_m}[1+\theta(1-3\eta^2)]}.
\end{equation}
This temperature is always higher than the corresponding temperature
with no cosmological constant. The maximum departure from the case with 
$\Lambda=0$ is 30\%, for $\theta=1/2$, for which $\eta=0.366$. 
Henceforth, we shall refer to the virial temperature in equation (\ref{temp_low}) 
as $T_{\rm SUS}$. For the case of a collisionless gas, we replace the virial 
temperature above by the virial velocity dispersion,
\begin{equation}
\label{dispersion_low}
\sigma_V^2=\frac{\langle v^2\rangle}3=\frac{k_BT}{m}.
\end{equation}

For a given $\theta$, the mean overdensity of the virialized object in
the SUS approximation with respect to the critical density of the universe 
at $z_{\rm coll}$ is
\begin{equation}
\label{Delta_c_SUS}
\Delta_c\equiv\frac{\bar{\rho}}{\rho_{\rm crit}(z_{\rm coll})}
	=\frac{\Omega_0a_0^3}{\theta\eta^3(\theta)}
		\displaystyle{\left[\frac{\rho_{\rm crit}(z_{\rm coll})}{\rho_{\rm crit,0}}\right]^{-1}},
\end{equation}
where $a_0$ is defined by equation (\ref{a0}), according to which $\Omega_0a_0^3=\lambda_0$ if
$\Omega_0+\lambda_0=1$, and where equations~(\ref{Friedmann}) and (\ref{z-a}) yield
\begin{equation}
\label{rho_crit_evol}
\displaystyle{\frac{\rho_{\rm crit}(z_{\rm coll})}{\rho_{\rm crit,0}}}
	=\displaystyle{\left[\frac{h(z_{\rm coll})}{h}\right]^{2}}
	=(1-\Omega_0-\lambda_0)(1+z_{\rm coll})^2+\Omega_0(1+z_{\rm coll})^3+\lambda_0.
\end{equation} 
Equations~(\ref{eta_SUS0}), (\ref{Delta_c_SUS}) and (\ref{rho_crit_evol}) form a set of 
simultaneous equations which may be solved for the dependence of $\eta$ and $\theta$
on $z_{\rm coll}$, once $\Delta_c$ is known by integrating the nonlinear top-hat
perturbation differential equation~(\ref{delta_gen_low}) [or (\ref{delta_open})]. 
In general, this solution for the dependence of $\eta$ and $\theta$ on $z_{\rm coll}$
for a given $\Delta_c$
involves the roots of a cubic equation which can be cumbersome to express analytically.
However, for $\theta<<1$ ($z_{\rm coll}>>1$), a good approximate solution is given by
\begin{equation}
\eta=2q+\frac12,
\end{equation}
and
\begin{equation}
\theta=\frac q{\eta^3},
\end{equation}
where
\begin{equation}
q\equiv\frac{\Omega_0a_0^3}{\Delta_c}\displaystyle{\left[\frac{h(z_{\rm coll})}{h}\right]^{-2}}.
\end{equation}
The dependence of the quantity $\Delta_c$ on $z_{\rm coll}$ can be expressed in terms 
of approximate analytical fitting formulae (with errors $\sim1\%$) according to
\begin{equation}
\label{Delta_SUS_approx}
\Delta_{ c}=18\pi^2+c_1x-c_2x^2,
\end{equation}
where $x\equiv\Omega(z_{\rm coll})-1$, and $c_1=82\, (60)$ and $c_2=39\,(32)$ for the
flat (open) cases, $\Omega_0+\lambda_0=1$ ($\Omega_0<1,\lambda_0=0$), respectively 
\cite{B&N}, where
\begin{equation}
\Omega(z)=\frac{\Omega_0(1+z)^3}{(1-\Omega_0-\lambda_0)(1+z)^2
+\Omega_0(1+z)^3+\lambda_0}.
\end{equation}
Henceforth, we shall refer to the quantities $\eta$ and $\Delta_c$ for the SUS 
approximation as $\eta_{\rm SUS}$ and $\Delta_{\rm c,SUS}$.
\section{Isothermal Spheres in the Presence of the X-component}
\label{isoth_X}
For matter-dominated universes, both open and flat,
 the final virialized object decouples from the 
expanding cosmological background from which it condensed.
Hence, when we describe it as an 
isothermal sphere in hydrostatic equilibrium, we do so in the usual 
non-cosmological way (e.g. Binney and Tremaine 1987).
Similarly, in the presence of
an X-component, the final virialized object also decouples from the 
expanding cosmological background from which it condensed, but 
it continues to be affected by the X-component because of the modification 
that component causes to the gravity force, as discussed earlier.
We can still describe it as an isothermal sphere in hydrostatic 
equilibrium in the usual non-cosmological way, but we must account properly
for this modification of the gravity force. The Poisson equation in
equation (\ref{Poisson}) for the gravitational potential in the case of 
spherical symmetry is 
\begin{equation}
\label{eqpot_sph_low}
\displaystyle{\frac 1{r^2} \frac d{dr} \left(r^2 \frac{d\Phi_{\rm tot}}{dr}\right)
	= -4 \pi G (\rho+\rho_{\rm X,eff})}.
\end{equation}
The equation of hydrostatic equilibrium, $\nabla p = \rho {\bf g}$,
 where $g=-\nabla\Phi_{\rm tot}$, combines with 
equation~(\ref{eqpot_sph_low}) in spherical symmetry to become
\begin{equation}
\label{hydrost_sph_low}
\displaystyle{\frac{k_BT}m \frac{d\rho}{dr} 
	= - \rho \nabla \Phi_{\rm tot} =-\rho \frac{GM(r)}{r^2}}
		-\rho\frac{4\pi Gr\rho_{\rm X,eff}}{3},
\end{equation}
where $M(r)$ is the mass inside radius $r$. 
Multiplying equation (\ref{hydrost_sph_low}) by $r^2m/\rho k_B T$ and taking the 
derivative with respect to $r$, we obtain
\begin{equation}
\label{eqrho_sph_low}
\displaystyle{\frac{d}{dr} \left(r^2 \frac{d(\ln \rho)}{dr}\right)
	= -4 \pi \frac{Gm}{k_B T}(\rho+\rho_{\rm X,eff})r^2}.
\end{equation}

Let us consider the case of collisionless particles, too. 
The equilibrium velocity distribution of the particles is a Maxwellian
distribution given by
\begin{equation}
f(v) = \displaystyle{\frac{\rho_0}{(2\pi \sigma_V^2)^{3/2}}
	\exp\left(\frac{\Phi_{\rm tot} - v^2/2}{\sigma_V^2}\right)},
\end{equation}
where $\rho_0$ is the central density if we take $\Phi_{\rm tot}(r=0)=0$, 
and $\sigma_V$ is the one-dimensional velocity dispersion.
After integrating over velocity, we obtain
\begin{equation}
\rho =\int f(v)d\,{\bf v}= \rho_0 e^{\Phi_{\rm tot}/\sigma_V^2},
\end{equation}
which we substitute into equation (\ref{eqpot_sph_low}), to obtain
\begin{equation}
\label{rhosig_sph_low}
\displaystyle{\frac d{dr}\left(r^2 \frac {d(\ln\rho)}{dr}\right)
	= -\frac{4\pi}{\sigma_V^2} G(\rho+\rho_{\rm X,eff}) r^2}.
\end{equation}
By calculating the mean square velocity we obtain
\begin{equation}
 \langle v^2\rangle=3\sigma_V^2.
\end{equation}
The equivalent temperature can be calculated from
\begin{equation}
\displaystyle{\frac{\langle v^2\rangle}{2}=\frac 32\frac{k_B T}{m}}
\end{equation}
to obtain for $\sigma_V$:
\begin{equation}
\label{sigma_sph_low}
\displaystyle{\sigma_V^2=\frac{k_BT}m}.
\end{equation}
A comparison of equation (\ref{rhosig_sph_low}) with equation (\ref{eqrho_sph_low}) 
using equation (\ref{sigma_sph_low}) shows they are identical.
Hence, the structure of a self-gravitating isothermal fluid sphere in
hydrostatic equilibrium for $\gamma=5/3$ is 
identical with that for a spherically-symmetric system of collisionless 
particles in virial equilibrium with three-dimensional particle orbits
which are isotropic and have a spatially uniform velocity dispersion.

To make equation (\ref{rhosig_sph_low}) nondimensional, we introduce new 
variables
\begin{equation}
\label{vars_low}
\tilde{\rho} = \displaystyle{\frac{\rho}{\rho_0}}, \qquad
\zeta = \displaystyle{\frac{r}{r_0}},\qquad
\tilde{\rho}_{\rm X,eff}= \displaystyle{\frac{\rho_{\rm X,eff}}{\rho_0}},
\end{equation}
where $\rho_0$ is the central density, and we choose
\begin{equation}
\label{r_0_low} 
r_0=\sigma_V/(4\pi G\rho_0)^{1/2}.
\end{equation}
 In terms of these variables, equation (\ref{rhosig_sph_low}) becomes
\begin{equation}
\label{nondim_sph_low}
\frac d{d\zeta}
	\left(\zeta^2 \frac{d(\ln\tilde{\rho})}{d\zeta}\right)
	= - \zeta^2(\tilde{\rho}+\tilde{\rho}_{\rm X,eff}).
\end{equation}
We must solve equation (\ref{nondim_sph_low})  with the following 
boundary conditions:
\begin{equation}
\label{init_cond_low}
\tilde{\rho}(0)=1, \qquad \displaystyle{\frac{d\tilde{\rho}}{d\zeta}(0)=0}.
\end{equation}

An important difference between equation (\ref{nondim_sph_low}) and the
standard isothermal Lane-Emden equation should be noted here. The 
standard form of the equation for the case without the X-component
does not provide a natural cutoff of the
density profile, and, thus, both the profile and the mass are infinite. In
the presence of an X-component, however, the modified Lane-Emden equation 
shows that the density profile is finite as long as
$n<2$, since there is a matter density below which the matter cannot 
be gravitationally bound. According to equation~(\ref{nondim_sph_low}), 
this cutoff occurs at 
$\rho(r)=-\rho_{\rm X,eff}$. For the special case of a cosmological constant, 
this cutoff occurs where $\rho=2\rho_\lambda$. The mass of any such halo is, 
therefore, also 
finite and depends on the value of $\rho_{\rm X,eff}$.

\begin{figure}
\centerline{\psfig{figure=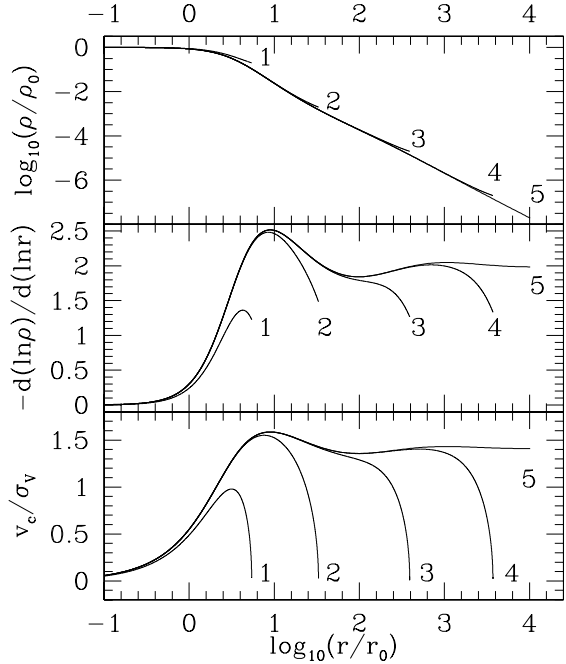,height=3.5in,width=3.5in}}
\caption{Modified Isothermal Lane-Emden Spheres in the Presence of 
a Cosmological Constant.
(top panel) Density profile (in units of the central density $\rho_0$)
versus radius (in units of the core radius $r_0$) when the X-component
is a cosmological constant, for $\tilde{\rho}_\lambda=0.1$ (label 1), 
$\tilde{\rho}_\lambda=0.001$ (label 2), $\tilde{\rho}_\lambda=10^{-5}$ (label 3), 
$\tilde{\rho}_\lambda=10^{-7}$ (label 4), and $\tilde{\rho}_\lambda=0$ 
(no cosmological constant; label 5).
Each plot extends out to its natural cutoff at $\rho=2\rho_{\lambda}$, except
the matter-dominated one (curve 5), which is infinite and is plotted only 
out to $r/r_0=10^4$. 
(middle panel) The logarithmic slopes of these density profiles versus 
dimensionless radius.
(bottom panel) The circular velocity profiles $v_c$ (normalized by the velocity 
dispersion of the halo $\sigma_V$) versus dimensionless radius. Note the steep 
fall of the circular velocity close to the cutoff radius due to the fact that
outer shells are only marginally
bound in the presence of cosmological constant, unlike the case $\Lambda=0$.}
\label{dens_low} 
\end{figure}

In Figure~\ref{dens_low}, we specialize to the cosmological constant case and 
plot the dimensionless density profile $\rho/\rho_0$, the logarithmic
slope of this density profile, and the dimensionless
circular velocity profile $v_c/\sigma_V$, normalized to the velocity dispersion 
$\sigma_V$, all as functions of $r/r_0$ for several representative values of 
$\tilde{\rho}_\lambda$. In the presence of a cosmological
constant, this circular velocity is:
\begin{equation}
\label{circ_vel}
\displaystyle{v_c^2(r)\equiv r\frac{d\Phi}{dr}
	=\frac{GM(r)}r\left(1-2\frac{\rho_{\lambda}}{\rho(r)}\right)}.
\end{equation}

 When $\rho_{\rm X,eff}=0$, there is a well-known, analytical, power-law 
solution to the isothermal Lane-Emden equation, if we relax the inner 
boundary conditions in equation (\ref{init_cond_low}) 
so as to permit a singularity at the origin. This
singular isothermal sphere (SIS) solution is given by
\begin{equation}
\label{rho_sing_low}
\rho(r)=\displaystyle{\frac{\sigma_V^2}{2\pi Gr^2}},
\end{equation}
and
\begin{equation}
\label{sigma_sing_low}
\sigma_V^2=\frac 12\frac{GM(r)}{r}={\rm constant}.
\end{equation}
For the modified isothermal Lane-Emden equation in the presence of the
X-component, however, no such analytical solution is possible. Nevertheless,
since $\rho(r)<<\rho_\lambda$ over most of the profile, except for the outer 
parts, the SIS solution is still a reasonable approximate solution of equation
(\ref{nondim_sph_low}) out to $\rho\sim\rho_\lambda$. As for the case
$\rho_{\rm X,eff}=0$, this approximate SIS solution when $\rho_{\rm X,eff}\neq0$
also does not satisfy the inner boundary condition in equations (\ref{init_cond_low}).
We will compare our results for the nonsingular case with this approximate
SIS solution in \S~\ref{compare_SUS_SIS_low} below,
after we complete our derivation of
the nonsingular TIS solution. 

As mentioned above, unlike the standard result when $\rho_{\rm X,eff}=0$,
isothermal spheres in the presence of the X-component always have a 
finite radius and
mass. Hence, unlike the case with no X-component, we do not need to truncate 
these spheres in order to apply them to describe realistic finite structures.
However, as will be shown below, our final TIS model
actually imposes a smaller truncation radius $r_t$ than that which corresponds to
the marginally bound mass shell in the isothermal sphere solution.
The total mass $M_0$ of the isothermal sphere is then
\begin{equation}
M_0=M(r_t)=\int^{r_t}_0 4\pi\rho(r) r^2d\,r=4\pi\rho_0r_0^3 \tilde{M}(\zeta_t),
\end{equation}
where $\zeta_t=r_t/r_0$ and $\tilde{M}(\zeta_t)$ is the dimensionless mass:
\begin{equation}
\label{mass_nond_low}
\tilde{M}(\zeta_t)\equiv
	\displaystyle{\frac {M(r_t)}{ 4 \pi r_0^3\rho_0}
	=\int^{\zeta_t}_0\tilde{\rho}(\zeta)\zeta^2 d\zeta}.
\end{equation}
If such a truncation radius $r_t$ does exist, 
this leads to the necessity of an external 
pressure to keep the system in equilibrium, which requires the inclusion
of a surface pressure term in the virial theorem. This correction and its 
consequences are discussed in the next section.

\section{The Virial Theorem for Truncated Isothermal Spheres in a
Low Density Universe}
\label{virial_low}

Let us consider the general isothermal sphere density profile, $\rho(r)$,
obtained in the previous section. From the ideal gas law, the pressure 
inside the sphere as a function of the radius is
\begin{equation}
p(r)=\frac{k_BT}m\rho(r)=\sigma_V^2 \rho(r),
\end{equation}
and at the outer edge
\begin{equation}
\label{p_t_low}
p_t= p(r_t)=\sigma_V^2 \rho(r_t).
\end{equation}
The mean pressure inside the sphere is
\begin{equation}
\bar{p}=\displaystyle{\frac{\int pd\,V}{\int d\,V}}
     =\displaystyle{\frac{3\int^{\zeta_t}_0\tilde{\rho}(\zeta)\zeta^2d\,\zeta}
		{\zeta_t^3\tilde{\rho}(\zeta_t)}p_t}
	=\displaystyle{\frac{3\tilde{M}(\zeta_t)}
		{\zeta_t^3\tilde{\rho}(\zeta_t)}p_t\equiv\alpha(\zeta_t)p_t},
\end{equation}
where $\zeta_t\equiv r_t/r_0$. 

The virial theorem for a static sphere in the presence of a surface pressure 
and of the X-component now becomes
\begin{equation}
\label{vir_th_low}
3(\gamma-1)K+W+S_p-2W_X=0,
\end{equation}
where K is once again just $U_{th}$, and $S_p$ is the surface pressure term.
The thermal energy for a gas with a ratio of specific heats $\gamma=5/3$ is 
given by
\begin{equation}
U_{th}=\frac 1{\gamma-1}\int pd\,V=\frac {\alpha(\zeta_t)p_tV}{\gamma-1}
	=\frac 32\alpha p_t V ,
\end{equation}
where $V$ is the total volume. The surface term is equal to
\begin{equation}
S_p=-\int p{\bf r}\cdot d\,{\bf S}=-3Vp_t.
\end{equation}
Hence, according to the virial theorem, the unmodified
gravitational potential energy term $W$ is
\begin{equation}
W=-2\frac{\alpha-1}\alpha U_{th}+2W_X,
\end{equation}
and the total energy is
\begin{equation}
\label{total_en_low}
E=\frac{2-\alpha}{\alpha}U_{th}+3W_X.
\end{equation}
In order for the halo to be bound, the condition $E<0$ must be 
met, which requires 
\begin{equation}
\label{bound}
\alpha>\frac{2}{1-3\frac{W_X}{U_{\rm th}}}.
\end{equation}
The virial temperature of this generalized isothermal sphere is 
\begin{equation}
\label{temper_low}
T=\frac {2\alpha}{5(\alpha-2)} \frac{G{M_0}m}{k_Br_m}[1+\theta(1-3\eta^2)],
\end{equation}
which we shall henceforth refer to as $T_{TIS}$\footnote{Henceforth, as 
in Paper I, the notation ``TIS'' shall refer to a solution of the 
isothermal Lane-Emden equation (modified here to account for 
the presence of the X-component), with the 
nonsingular boundary conditions of equation (\ref{init_cond_low})
at the origin.}. 

For $\theta = 0$, the temperature in equation (\ref{temper_low}) reduces to
$T_{\rm TIS}$ derived in Paper I. In the presence of the X-component, this 
temperature is always slightly higher than the corresponding temperature 
for $\rho_X=0$, however, but never by more than 5\%, with the maximum 
reached for  $\theta=0.5$.
Since $\alpha/(\alpha-2)>1$ for any $\alpha$, the 
temperature $T_{TIS}$ is always higher than $T_{SUS}$, the standard value for
a uniform sphere shown in equation (\ref{temp_low}). 

Just as we did in Paper I, we identify the virial radius $r_{\rm vir}$ 
with the size of the truncated isothermal sphere (i.e. $r_{\rm vir}=r_t$). 
For comparisons with the results of 
dynamical calculations of the formation of such an equilibrium object, we 
should interpret $r_{\rm vir}$ (and $r_t$) as the radius inside which
hydrostatic equilibrium holds (e.g. Cole \& Lacey 1996). Hereafter, we specialize the
discussion to the case in which the X-component is the cosmological constant. 

\section{Choosing a Unique Profile: The Minimum-Energy Solution 
in a Low-Density Universe}
\label{min_en} 
\subsection{The Nonsingular, Truncated Isothermal Sphere Formed
by Top-Hat Collapse}
\label{TIS_low}
The virialized object which results from the
collapse of a given top-hat density perturbation must have the mass of the 
top-hat before it collapsed and virialized. We assume that the total energy 
$E$ is also conserved, including the effect the cosmological constant 
has on the potential energy, if
present. Fixing the mass $M_0$ and the energy $E$ of the TIS which
results from the collapse of the top-hat, however, is not enough to specify 
which member of the infinite family of solutions of the modified isothermal
Lane-Emden equation which all share this mass and energy is chosen 
by the collapsing top-hat. In order to make a unique choice, some additional 
information is required, such as the value of the boundary pressure $p_t$. As 
in Paper I, we shall assume that the correct boundary
pressure $p_t$ is that value which makes the TIS with this total energy $E$
and mass $M_0$ correspond to the unique minimum-energy solution.
In order to determine the minimum-energy TIS which results from a given 
top-hat collapse, we must take account of the fact that this top-hat collapse 
depends upon the value of the parameter $\theta$. This parameter measures 
the relative importance of the vacuum energy density associated with the 
cosmological constant and the rest-mass energy density of the matter 
inside the top-hat at maximum expansion, which is different for different 
epochs of maximum expansion even for a given value of the cosmological constant.
 As a result, the internal structure of the TIS solution which results from
a given top-hat collapse even when expressed in dimensionless form with
radius in units of $r_0$ and density in units of $\rho_0$ will depend, 
not only upon the background cosmology parameters (i.e. $\Omega_0=1-\lambda_0$
and $H_0$), but also upon $z_{\rm coll}$ for the top-hat. This makes the 
calculation of our minimum-energy TIS solution in this case more complicated
than for the EdS case in Paper I.  


As in Paper I, this conservation of the energy $E$ of the top-hat before and after its 
collapse and virialization assumes that there is no extra $pdV$-work that 
needs to be taken into account due to the presence of the external boundary 
pressure $p_t$ which would otherwise alter the final total energy $E$ 
compared with its initial value before collapse. This is appropriate for 
the case at hand, since the collapse prior to the epoch of virialization 
is that of a cold, pressure-free gas which collapses supersonically. As 
in the well-known, self-similar, spherical infall solution of
Bertschinger (1985), the original energy is converted from potential energy at
maximum expansion to a mixture of infall kinetic energy and potential energy 
during infall, with a negligible share of the energy going into compressional 
heating. In that solution, the infall is halted by a strong shock. At this 
shock, the kinetic energy of infall is converted primarily into the thermal 
energy of the shock-compressed gas, with only a small portion remaining as 
kinetic energy of subsonic, postshock infall. While the self-similar infall 
solution is strictly correct only for the EdS universe, we expect similar 
behaviour in low density universes, as well. However, since no general self-similar
infall solution exists in these cases, the analogous infall solution for
low-density universes must be studied by numerical means. Of course, in the limit
of high redshift, the self-similar solution of the EdS case can be applied even to 
the case of a low-density universe if properly rescaled (see Appendix~\ref{appendix1}).

By analogy with this infall solution, therefore, we
identify the boundary pressure $p_t$ in the case of our TIS solution, not as a
fixed external pressure which acts on the top-hat boundary throughout its collapse 
and virialization, but rather as something like the instantaneous post-shock 
pressure in the infall solution. As such, it has a physical origin in the 
conversion of the original energy of the collapsing top-hat, itself, from 
potential energy at maximum expansion into kinetic energy of infall during 
collapse and, finally, into thermal energy of the post-shock gas, always 
conserving the original energy $E$ of the top-hat.  
 
\subsection{Finding the Unique Minimum-Energy Solution}
\label{min_en_low}
\subsubsection{Flat Universe with Cosmological Constant}
\label{min_flat}
As shown above, the truncated isothermal sphere solutions form a one-parameter
family, described by $\zeta_t=r_t/r_0$ -- the truncation radius in units of 
the core radius. Specifying $\zeta_t$, the total mass, and the total energy 
completely determines the solution. Alternatively, we can specify 
the mass, the total energy and the applied
external pressure $p_t$. 

\begin{figure}
\centerline{\psfig{figure=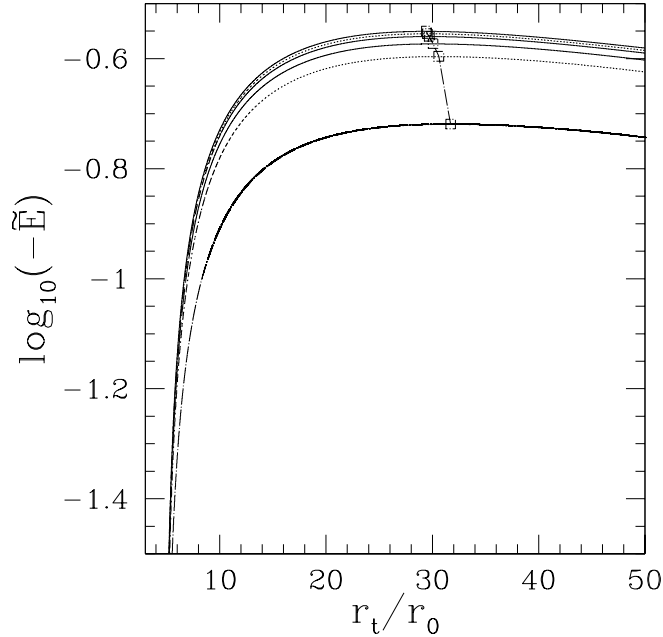,width=3.5in}}
\caption{ 
The dimensionless energy $\tilde{E}(\zeta_t)$ versus dimensionless 
truncation radius $\zeta_t=r_t/r_0$ for several representative
values of $\theta$ (see also Table~\ref{summary_low}), chosen as described in 
\S\ref{summary_tis_low}. From top to bottom:
$\theta=0$ (matter-dominated models), $\theta=0.0118$, 
$\theta=0.0249$, $\theta=0.0604$, $\theta=0.123$, and $\theta=0.5$. The symbols 
indicate the minimum-energy points for each curve.}
\label{energy_plot_low}
\end{figure}

The conservation of energy equates the energy after collapse and virialization 
to the energy of the top-hat at maximum expansion, $E(t_m)=E_{\rm vir}$. This
 yields
\begin{equation}
\label{en_conserv}
\displaystyle{-\frac35\frac{GM_0^2}{r_m}\left(1+\frac{\rho_\lambda}{\rho_m}\right)
	=\frac{2-\alpha}{\alpha}U_{\rm th}M_0-16\pi^2G\rho_\lambda I_t,}
\end{equation}
where
\begin{equation}
\label{I}
I_t\equiv \int_0^{r_t}\rho r^4dr.
\end{equation}
In dimensionless variables, equation (\ref{en_conserv}) becomes
\begin{equation}
\label{en_conserv_nond}
\displaystyle{(1+\theta)\eta_{\rm TIS}=\frac52\frac{\alpha-2}{\alpha}\frac{\zeta_t}{\tilde{M}_t}
	+\frac{5\theta\eta_{\rm TIS}^3\tilde{I}_t}{\tilde{M}_t\zeta^2_t}},
\end{equation}
where
\begin{equation}
\label{I_nond}
\tilde{I}_t\equiv \int_0^{\zeta_t}\tilde{\rho} \zeta^4d\zeta,
\end{equation}
and we have used the fact that
\begin{equation}
\label{rho_lam}
\displaystyle{\tilde{\rho}_{\lambda}=\frac{\rho_\lambda}{\rho_0}
	=\frac{\rho_\lambda}{\rho_m}\frac{\rho_m}{\rho_0}
	=\frac{3\theta\eta_{\rm TIS}^3\tilde{M}_t}{\zeta^3_t}.}
\end{equation}
For a given external pressure $p_t$, the total energy of the modified
isothermal Lane-Emden sphere is given by
\begin{eqnarray}
\label{tot_en}
E&=&\displaystyle{-\frac 35\frac{GM_0^2}{r_m}(1+\theta)}
=\displaystyle{-\frac 35\frac{GM_0^2}{r_0}(1+\theta)\frac{\eta_{\rm TIS}}{\zeta_t}}
				\nonumber\\
&=&\displaystyle{\frac35(4\pi^3G)^{1/4}M_0^{3/2}p_t^{1/4}(1+\theta)\tilde{E}(\zeta_t)},
\end{eqnarray}
where
\begin{equation}
\label{E_nond}
\displaystyle{\tilde{E}(\zeta_t)
	=-\frac{\eta_{\rm TIS}\tilde{M}_t^{1/2}}{\zeta_t\tilde{\rho}_t^{1/4}}}.
\end{equation}
We have plotted the dependence of the dimensionless total energy $\tilde{E}$
on $\zeta_t$ in Figure \ref{energy_plot_low} for several representative values of
the parameter $\theta$. 
In order to indicate the 
dependence of the size of the sphere on $\zeta_t$, we can nondimensionalize the 
radius $r_t$, according to 
\begin{equation}
\label{lambda_E_low}
-\lambda_E\equiv\displaystyle{-\frac{r_tE}{GM_0^2}
	=\frac 35(1+\theta)\eta_{\rm TIS}}.
\end{equation}
This definition of $\lambda_E$ also corresponds to the familiar 
dimensionless energy
parameter used in discussions of the stability of isothermal spheres.

For a given mass $M_0$ and external pressure $p_t$, the solution is specified
uniquely only if we can uniquely identify a special value of $\zeta_t$, or
equivalently, of $E$.
Apparently, for any truncated isothermal sphere of mass $M_0$ which is confined
by a given external pressure $p_t$, there is a unique value of $\zeta_t$ which
minimizes the total energy $E$. As in Paper I, we shall make the 
reasonable anzatz that this 
minimum-energy solution is the unique TIS solution preferred in nature as the 
outcome of the virialization of the sphere in the presence of a fixed external 
pressure. We offered evidence to support this anzatz in Paper I.
It is possible that this is a general result for any TIS formed by 
relaxation in the presence of a fixed external pressure, but we are 
only concerned here with spheres that evolve from cosmological initial 
conditions. 

\begin{figure}
\centerline{\psfig{figure=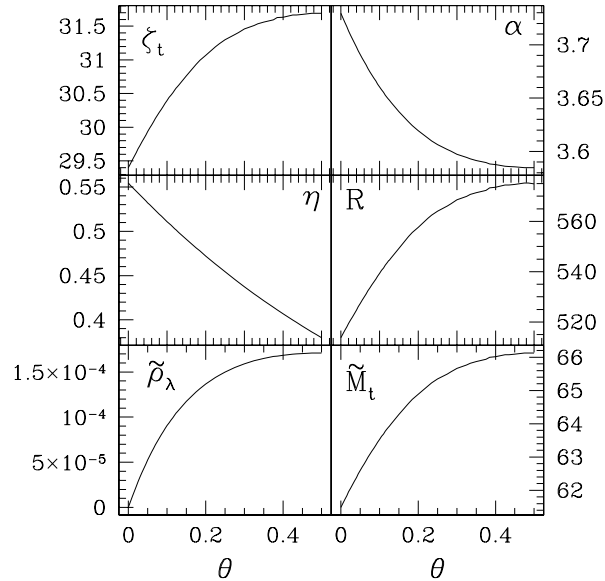,width=0.55\textwidth}}
\caption{Dimensionless parameters of the TIS model in a flat universe with
cosmological constant vs. the parameter $\theta$, where $\theta\equiv\rho_\lambda/\rho_m$,
the ratio of the vacuum energy density to the matter rest-mass energy density at
maximum expansion of the parent top-hat.  
}
\label{6panel}
\end{figure}

For each value of $\theta$ in the allowed range $0\leq\theta\leq0.5$,
the minimum value of $E$ as a function of $\zeta_t$ for a given $p_t$ is found 
by minimizing the dimensionless energy $\tilde{E}(\zeta_t)$ in equation 
(\ref{E_nond}), with $\eta_{\rm TIS}$ found by solving 
equation~(\ref{en_conserv_nond}). 
Unlike the EdS case discussed in Paper I,
the minimum-energy solution in a low-density universe is not universal 
(except for the special case $\Lambda=0$), but
instead depends upon the time of collapse, as implicitly parameterized by
the quantity $\theta$. This makes our calculation of the solution
more complicated than in Paper I, since equation~(\ref{nondim_sph_low}) depends upon 
$\tilde{\rho}_\lambda$ and must, therefore,
 be solved in conjunction with equation~(\ref{rho_lam}) to obtain a 
self-consistent solution and, subsequently, to find the minimum-energy point.
As can be seen in Figure~\ref{energy_plot_low}, the minimum of the energy
corresponds in each curve to a unique value of  the dimensionless truncation
radius $\zeta_t$ and becomes more pronounced when the cosmological constant 
contribution is more important. The value of $\zeta_t$ at which this minimum 
energy occurs increases with increasing 
$\tilde{\rho}_\lambda$, from $\zeta_t=29.4$ for matter-dominated 
models ($\theta=0$) to $\zeta_t=31.7$ for $\theta=0.5$, the maximum allowed value. 
The resulting values for  
$\zeta_t$, $\alpha(\zeta_t)$, $\eta_{\rm TIS}$, $R\equiv\rho_0/\rho_t$,
$\tilde{\rho}_\lambda$, and $\tilde{M}_t$ versus $\theta$ are shown in 
Figure~\ref{6panel}. The point $\theta=0$ corresponds to the solution for 
matter-dominated models without cosmological constant (either open or flat), 
where these dimensionless quantities are identical to those for the EdS case
obtained in Paper I. 

Thus far, we have derived the unique, dimensionless minimum-energy
TIS solution for the postcollapse virialized object which results from top-hat 
collapse at any epoch for different cosmologies, as parameterized by the value of 
$\theta=\rho_\lambda/\rho_m$, the ratio of the vacuum energy density to the matter 
rest-mass energy density of the top-hat at its epoch of maximum expansion. 
In order to complete the dimensionless description of the TIS model for the 
postcollapse equilibrium object created by a top-hat which 
collapses at a given redshift $z_{\rm coll}$, it is necessary to relate $\theta$ to 
$z_{\rm coll}$, according to the solution of the nonlinear top-hat perturbation 
equation~(\ref{delta_gen_low}) [or (\ref{delta_open})]. This dependence of $\theta$
on $z_{\rm coll}$ for the top-hat perturbation was described in \S~\ref{top_low_dens} 
above.

The parameters we have derived for the unique minimum-energy TIS solution for several 
representative values of the parameter $\theta$, chosen so as to span the allowed 
range, are summarized in Table~\ref{summary_low}. These values of the parameter $\theta$
were chosen to correspond to the EdS case and open-matter-dominated cases ($\theta=0$), 
and to the flat, $\Lambda$-dominated model currently favoured by astronomical
observations, with $\lambda_0=0.7$, for different
collapse epochs: $z_{\rm coll}=1$ ($\theta=0.01183$), 
$z_{\rm coll}=0.5$ ($\theta=0.0249$), and $z_{\rm coll}=0$ ($\theta=0.0604$), as
well as for the extreme cases of $\lambda_0=0.9$, $z_{\rm coll}=0$ ($\theta=0.123$), and
of either $\lambda_0\rightarrow1$ or else $1+z_{\rm coll}\rightarrow0$ (i.e. in the
future) ($\theta=0.5$).
If we adopt $\lambda_0\leq0.9$ as a reasonably conservative estimate of the
possible range allowed  by astronomical observation, then the maximum value of
$\theta$ of interest corresponds to that for which $z_{\rm coll}=0$, i.e. 
$\theta(\lambda_0=0.9,z_{\rm coll}=0)=0.123$.

The radius $r_t$ of the TIS solution is encountered by the outer boundary of 
the collapsing top-hat at a time somewhat earlier than the time $t_{\rm coll}$ of 
infinite density in the top-hat solution. We shall refer to the time at 
which the radius of the collapsing top-hat equals $r_t$ as $t_{\rm cross}$, 
to distinguish it from $t_{\rm coll}$.  
At time $t_{\rm cross}$, the top-hat exact solution yields an overdensity 
$\delta=\delta_{\rm cross}\approx 100$ for EdS universe, while the extrapolated 
linear solution at this time yields $\delta_{L,cross}\approx 1.56$. For a flat,
low-density universe, however, both $\delta_{\rm cross}$ and $\delta_{\rm L,cross}$ 
exceed their values for the EdS case. For example, if $z_{\rm coll}=0$, then 
$\delta_{\rm cross}=197 (416)$ and $\delta_{\rm L, cross}=1.96 (2.52)$ for 
$\lambda_0=1-\Omega_0=0.7 (0.9)$, respectively. For $\lambda_0\rightarrow1$ or 
else in the future for any value of $\lambda_0$, both 
$\delta_{\rm cross}\rightarrow\infty$ and $\delta_{\rm L, cross}\rightarrow\infty$.
If we assume that
the TIS forms instantaneously at $t=t_{\rm cross}$, then its mean density
corresponds to a mean overdensity $\bar{\delta}={\delta}_{\rm cross}$ 
when compared with the background density at $z_{\rm cross}$, but the same TIS
corresponds to a mean overdensity $\bar{\delta}={\delta}_{\rm coll}$, which is
larger than this ${\delta}_{\rm cross}$ by about 30\%, when 
compared with the background density at 
$z_{\rm coll}$. This value of $\delta_{\rm coll}$
differs somewhat from the conventional value of $\bar{\delta}(t_{\rm coll})$ 
found for the postcollapse virialized sphere in the SUS approximation. 
The values of ${\delta}_{\rm cross}$, $\delta_{L,cross}$, and 
${\delta}_{\rm coll}$ for our chosen set of illustrative values of $\theta$ 
are also given in Table~\ref{summary_low}.  

As noted in Paper I, applications of the top-hat model involving the 
Press-Schechter approximation customarily identify the characteristic 
time of formation of objects of a given mass as the finite time at which the 
nonlinear perturbation solution predicts
collapse to infinite density, at which time the linear solution
yields $\delta_L=\delta_{\rm crit}$. 
Since $\delta_{L,cross}<\delta_{\rm crit}$, it may be appropriate to replace 
$\delta_{\rm crit}$ in such
applications by the value $\delta_{L,cross}$, implying that the 
number of objects formed at any epoch with a mass greater than some value may
be somewhat higher for the TIS solution than previously assumed in applications
of the Press-Schechter approximation.

\subsubsection{Open, Matter-Dominated Universe: Low-Density Universe without X-component}
For a matter-dominated universe, $\theta=0$, and the family
of solutions obtained in \S~\ref{min_flat} reduces to a single dimensionless
solution, which is independent of $\Omega_0$, identical to the 
solution we obtained in Paper I. The dimensional solution, however, 
depends upon the background cosmology. The details of this dependence will be 
discussed in \S~\ref{summary_tis_low}. As for the case of  the flat,
low-density universe discussed above, both $\delta_{\rm cross}$ and $\delta_{\rm L,cross}$ 
for the low-density, matter-dominated case exceed their values for the EdS case. 
For example, if $z_{\rm coll}=0$, then $\delta_{\rm cross}=222$ and 
$\delta_{\rm L, cross}=3.29$ for 
$\lambda_0=0,\Omega_0=0.3$. As $\Omega(z)\rightarrow0$ (i.e. in the future), we again find that
both $\delta_{\rm cross}\rightarrow\infty$ and $\delta_{\rm L, cross}\rightarrow\infty$.

\subsubsection{Stability}

The TIS model in an EdS universe represents a stable solution of the isothermal 
Lane-Emden equation (Paper I). This conclusion holds for any matter-dominated 
model, as well. In the presence 
of an X-component, however, the Lane-Emden equation is modified, leading to
a case whose stability, as far as we know, has not been studied. Nevertheless, 
since the TIS solution in the presence of the X-component departs from the 
solution for a matter-dominated universe by only a small amount for the 
observationally-constrained values of $\lambda_0$, we expect our 
solution for the low-density universe cases considered here to be stable, as well. 
However, detailed study of this problem is beyond the scope of this paper.

\subsection{Dependence of the TIS Model Parameters on Halo Mass and Collapse 
Redshift for Different Background Cosmologies}
\label{summary_tis_low}

In this section, we describe how the dimensional parameters
for our dimensionless TIS solution are specified for a given mass $M_0$ and
collapse redshift $z_{\rm coll}$.
For a given $\theta$, the dimensionless minimum-energy TIS solution and the 
nonlinear top-hat perturbation solution combine to yield the mean overdensity
of the TIS solution with respect to the critical density of the universe at
$z_{\rm coll}$,
\begin{equation}
\label{Delta_c_TIS}
\Delta_{\rm c,TIS}\equiv\frac{\bar{\rho}}{\rho_{\rm crit}(z_{\rm coll})}
	=\frac{\Omega_0a_0^3}{\theta\eta_{\rm TIS}^3(\theta)}
		\displaystyle{\left[\frac{\rho_{\rm crit}(z_{\rm coll})}{\rho_{\rm crit,0}}\right]^{-1}},
\end{equation}
(where $\Omega_0a_0^3=\lambda_0$ for $\Omega_0+\lambda_0=1$).
 Note that this overdensity is somewhat different 
from the corresponding overdensity $\Delta_{\rm c,SUS}$ derived from the standard 
uniform sphere approximation as described in \S~\ref{SUS}, 
since the collapse factors 
$\eta=r_m/r_t$ are different for the two cases. These two overdensities are 
simply related, however, as follows:
\begin{equation}
\label{Delta_c_rel}
\Delta_{\rm c,TIS}=\frac{\eta_{\rm SUS}^3}{\eta_{\rm TIS}^3}\Delta_{\rm c,SUS}.
\end{equation}
The dimensional parameters for the TIS solution for a  given
 total halo mass $M_0$ and $\Delta_{\rm c,TIS}$ can be expressed as follows:
\begin{eqnarray}
\label{r_m}
r_t &=&\displaystyle{\left(\frac{3M_0}
		{4\pi\Delta_{\rm c,TIS}\rho_{\rm crit}(z_{\rm coll})}\right)^{1/3}},\\
r_m &=& \frac{r_t}{\eta_{\rm TIS}}=\frac{\zeta_t}{\eta_{\rm TIS}}r_0,\\
\rho_0&=&\frac{\zeta_t^3}{3\tilde{M}_t}\bar{\rho}=\frac{\zeta_t^3}{3\tilde{M}_t}\Delta_{\rm c,TIS}
		\rho_{\rm crit}(z_{\rm coll}),\\
\sigma_V^2&=&4\pi G\rho_0r_0^2
	=\displaystyle{\left[\frac{4\pi}3\Delta_{\rm c,TIS}\rho_{\rm crit}(z_{\rm coll})\right]^{1/3}
	\frac{\zeta_t}{\tilde{M}_t}GM_0^{2/3}},\\
T_{\rm TIS} &=&\frac{m}{k_B}\sigma_V^2,\\
v_c(r)&=&
   \displaystyle{\left[\frac{GM(r)}{r}\left(1-2\frac{\rho_{\lambda}}{\rho(r)}\right)\right]^{1/2}}.
\label{v_c}
\end{eqnarray}
%
%
\begin{table*}
\begin{minipage}{170mm}
\caption{Summary of the minimum-energy TIS solution in a flat universe with a cosmological 
constant for several illustrative values of $\theta$.}
\label{summary_low}
\begin{tabular}{@{}lllllll}
Quantity&               $\theta=0$ &$\theta=0.0118$ &$\theta=0.0249$ &$\theta=0.0604$ &$\theta=0.123$ &$\theta=0.5$ \\ \hline
$\zeta_t=\frac{r_t}{r_0}$        & 29.400   & 29.534  & 29.677  &  30.04      & 30.58          & 31.69        \\[3mm]
$\alpha(\zeta_t)=\frac{\bar{p}}{p_t}$ & 3.730    & 3.720   & 3.709   &  3.684      & 3.649           & 3.585        \\[3mm]
$b_T=\frac{T_{TIS}}{T_{SUS}}$         & 2.156    & 2.163   & 2.170   &  2.188      & 2.213           & 2.262        \\[3mm]
$\tilde{M}(\zeta_t)$           & 61.48    & 61.76   & 62.06   & 62.80       & 63.90           & 66.13        \\[3mm]
$\tilde{\rho}_{\lambda}$         & 0  &$1.405\times10^{-5}$  & $2.842\times10^{-5}$ & $6.166\times10^{-5}$   &$1.042\times10^{-4}$&$1.710\times10^{-4}$\\[3mm]
$\eta_{\rm TIS}=\frac{r_t}{r_m}$           & 0.5544   & 0.5490  & 0.5432  & 0.5277      & 0.5016          & 0.3800       \\[3mm]
$\eta_{\rm SUS}$            & 0.5      & 0.4970  & 0.4937  & 0.4845      & 0.4676          & 0.3660       \\[3mm]
$\frac{r_{t,\rm TIS}}{r_{t,\rm SUS}}
		$    & 1.109 & 1.105 & 1.100 & 1.089 & 1.073       & 1.038        \\[3mm]
$\delta_{L,{\rm cross}}$         &1.562\footnote{These values are slightly different from
the values quoted in Paper I, Table 1. The correct values are shown here.
If we match the outer radius of this top-hat at its turnaround epoch to
the radius of the mass shell in the self-similar infall solution
of Bertschinger (1985) which contains the same mass at the same turnaround epoch,
which we did in Paper I, that mass shell crosses the shock at $\delta_L=1.5572$.}
					              & 1.630 & 1.710 & 1.955 & 2.524 & $\infty$\\[3mm]
$\delta(t_{\rm cross})$           
					  & $102.6^a$ & 116.2 & 133.5 & 196.8   & 416.1 & $\infty$\\[3mm]
$\delta_{\rm TIS}(t_{\rm coll})$           & $129.6$     & 148.5 & 172.7 & 264.4   & 579.1 & $\infty$\\[3mm]
$\frac{\rho_t}{\rho_0}$ &$1.946\times 10^{-3}$&$1.934\times10^{-3}$
		            & $1.920\times10^{-3}$ & $1.887\times10^{-3}$ & $1.837\times10^{-3}$ & $1.739\times10^{-3}$\\[3mm]
${\cal R}$              & 514 & 517 & 521 & 530 & 544 & 575\\[3mm]
$\frac{\rho_0}{\rho_{\rm b,coll}}$
    & $1.796\times 10^4$ & $2.064\times 10^4 $ & $2.430\times 10^4 $ & $3.784\times10^4$& $8.641\times 10^4 $ &$\infty$\\[3mm]
$\frac{\rho_t}{\rho_{\rm b,coll}}$
		&35.0& 39.92 & 46.67 & 71.39 & 158.8 & $\infty$ \\[3mm]
$\displaystyle{\lambda_E}$
		& -0.3326 & -0.3333 & -0.3340 & -0.3357 & -0.3380 & -0.3420\\ 
$K/|W|$&0.6832&0.6846&0.6861&0.6897 & 0.6948 & 0.7066\\ \hline
\end{tabular}
\end{minipage}
\end{table*}

For practical application of the TIS solution, we have provided a convenient
set of accurate analytical fitting formulae for the dependences of the 
dimensionless parameters of the solution (e.g. $\zeta_t$, $\eta_{\rm TIS}$, 
$\tilde{M}_t$) on $\theta$ in Appendix~\ref{appendix2}. Since fitting formulae are also 
available for the dependence of $\Delta_{\rm c,SUS}$ on $z_{\rm coll}$ for 
different cosmological parameters $(\lambda_0,\Omega_0)$, in the form of 
equation~(\ref{Delta_SUS_approx}), it is useful to express the dependence of
$\theta$ on $z_{\rm coll}$ for a given background cosmology by combining 
equations~(\ref{Delta_c_SUS}) and (\ref{rho_crit_evol}) to write
\begin{equation}
\label{theta}
\theta=\frac{\Omega_0a_0^3}{\Delta_{\rm c,SUS}\eta_{\rm SUS}^3}
		\displaystyle{\left[\frac{h(z_{\rm coll})}{h}\right]^{-2}}.
\end{equation}
In that case, for a given $z_{\rm coll}$, $\Omega_0$, $\lambda_0$, and $h$,
equation~(\ref{theta}) combines with equation~(\ref{eta_SUS}) which relates
$\eta_{\rm SUS}$ to $\theta$ and the fitting formulae for $\Delta_{\rm c,SUS}$
as a function of $z_{\rm coll}$ in equation~(\ref{Delta_SUS_approx}) to yield a 
pair of simultaneous algebraic equations for $\theta$.

\begin{figure}
\centerline{\psfig{figure=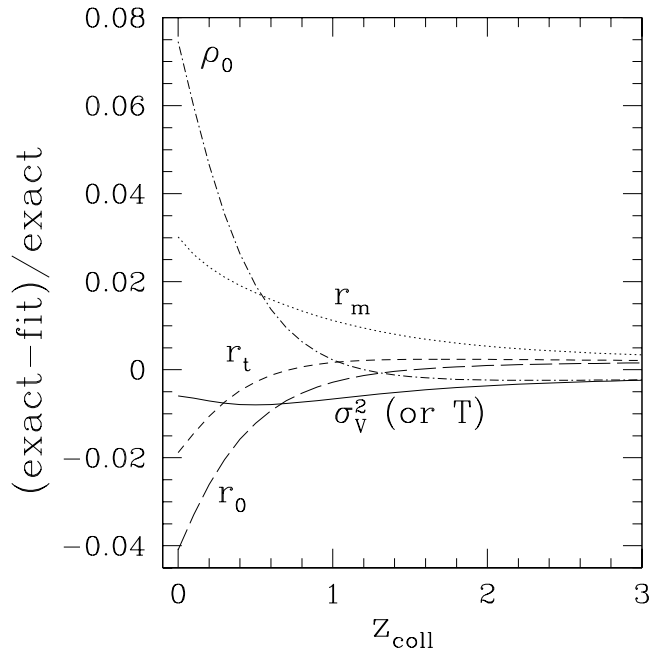,height=3.5in,width=3.5in}}
\caption{Fractional deviation of the approximate analytical fits in equations
(\ref{r_m_approx}) -- (\ref{rho0_approx}) from the 
exact numerical solution
for background cosmology with
$\lambda_0=1-\Omega_0=0.7$, as labelled.}
\label{errors}
\end{figure}

For most cases of current interest (i.e. $\Omega_0\geq0.3$, which corresponds 
to $\theta\leq0.06$ for $z_{\rm coll}\geq0$), a further approximation to these 
more accurate results is possible which is even simpler to use,
 if we neglect the effect of the cosmological constant on the
dimensionless solution of the Lane-Emden equation. In particular, as long as we 
take proper account of the evolution of a top-hat density perturbation in a low-density
universe up to its moment of infinite collapse and still satisfy energy conservation 
and the virial theorem, we will capture most of the dependence of the TIS solution
on the background cosmology. The result can be expressed as follows:
\begin{eqnarray}
\label{r_m_approx}
r_m &=& 
   337.7 \left(\frac {M_0}{10^{12}M_{\odot}}\right)^{1/3}[F(z_{\rm coll})]^{-1}
		h^{-2/3}\,\, {\rm kpc},\\ 
r_t &=&  187.2 \left(\frac {M_0}{10^{12}M_{\odot}}\right)^{1/3}
	[F(z_{\rm coll})]^{-1} h^{-2/3}\,\, {\rm kpc},\\
r_0&=&
	 6.367\left(\frac {M_0}{10^{12}M_{\odot}}\right)^{1/3}[F(z_{\rm coll})]^{-1}
		h^{-2/3}\,\, {\rm kpc},\\
T &=&
	 7.843\times 10^5\displaystyle{\left(\frac{\mu}{0.59}\right)\left(\frac {M_0}{10^{12}M_{\odot}}\right)^{2/3}
	F(z_{\rm coll})h^{2/3}\,\, {\rm K},}\\
\sigma_V^2 &=&
	1.098\times10^4\left(\frac {M_0}{10^{12}M_{\odot}}\right)^{2/3}
	F(z_{\rm coll})h^{2/3}\,\, {\rm km^2\,s^{-2}},\\
\label{rho0_approx}
\rho_0&=& 
	1.799\times10^4[F(z_{\rm coll})]^3\rho_{b0}
	=3.382\times 10^{-25}[F(z_{\rm coll})]^3h^2 \,\,{\rm g/cm^3}.
\end{eqnarray}
where 
\begin{equation}
F(z_{\rm coll})\equiv\displaystyle{\left[\frac{h(z_{\rm coll})}{h}\right]^2
\frac{\Delta_{\rm c,TIS}(z_{\rm coll},\lambda_0)}{\Delta_{\rm c,TIS}(\lambda_0=0)}}
=\displaystyle{\left[\frac{\Omega_0}{\Omega(z_{\rm coll})}\frac{\Delta_{\rm c,SUS}}{18\pi^2}\right]^{1/3}}
(1+z_{\rm coll}). 
\end{equation}
For the EdS case, $F=(1+z_{\rm coll})$, while for an open, matter-dominated universe
and a flat universe with a cosmological constant, $F\rightarrow\Omega_0^{-1/3}(1+z_{\rm coll})$ 
at early times [i.e. $x\rightarrow0$].
Here $\mu$ is the mean molecular weight, where
$\mu= 0.59\, (1.22)$ for an ionized (neutral) gas of H and He with 
$[He]/[H]=0.08$ by number. The relative accuracy of these formulae in comparison with
the exact numerical results is plotted  for the case $\lambda_0=1-\Omega_0=0.7$ in 
Figure~\ref{errors}, which shows that the errors are always small, even for 
$z_{\rm coll}=0$. 

\section{summary and discussion}
\label{summary}

\subsection{Comparison of the Minimum-Energy TIS Solution with the Standard Uniform
Sphere and Singular Isothermal Sphere Approximations}
\label{compare_SUS_SIS_low}

The temperature derived here for the minimum-energy
TIS solution for the postcollapse sphere in virial and hydrostatic
equilibrium which follows top-hat collapse is a factor of approximately two 
larger than the value previously derived for the SUS approximation, i.e.
by satisfying energy conservation and the virial theorem 
for a postcollapse sphere of uniform density for which the surface pressure 
term is neglected, as summarized in \S~\ref{top_low_dens}. In particular, if we write
$T_{TIS}=b_T T_{SUS}$, then $b_T=\alpha/(\alpha-2)\approx 2.2$. 
The size of the TIS sphere, given in terms of the radius
$r_m$ of the top-hat at maximum expansion according to $r_t=\eta_{\rm TIS} r_m$, is
actually not far from the size of the SUS sphere, $r_{\rm vir}=\eta_{\rm SUS} r_m$, as
shown in Table~\ref{summary_low}. However, in
terms of the TIS core radius $r_0$ (where, recall, our $r_{0,\rm TIS}\equiv r_{\rm King}/3$
as defined for conventional isothermal Lane-Emden spheres), the truncation radius is large, 
$r_t/r_0\geq29.4$, implying that the TIS is very far from uniform density.   
Despite this relatively small core radius, the TIS 
solution is also quite different from that of a {\it singular}
 isothermal sphere\footnote{As discussed in \S~\ref{isoth_X}, the singular
isothermal sphere is not an exact solution of the Lane-Emden equation in the 
presence of a cosmological constant, but for the purpose of this comparison, we 
ignore the effect of $\Lambda\neq0$ on this equation but take account of the more 
important effects of $\Lambda\neq0$ on the top-hat evolution, on the conservation of 
energy of the top-hat before and after its collapse and virialization and on the virial 
theorem.}.For the latter, the ratio of the average
density to that at the surface is 3, while we find a value of 3.6 to 3.7 for the
nonsingular TIS. The truncation radius in the singular limit is only $r_t=(5/12)r_m$,
smaller than the radius of the TIS.  
Similarly, the correction factor for the virial temperature of this SIS
relative to that of the SUS is $b_T=3$, as 
opposed to the value for the TIS model
which is $b_T\approx2$.  The central density
of the TIS is $>1.8\times 10^{4}$ times larger than the mean density
of the background at the collapse epoch for the parent top-hat.  Finally,
the TIS solution predicts that the top-hat will virialize somewhat earlier 
than the nominal collapse time of the top-hat,
since the outermost mass elements encounter a shock and shell-crossings
in the infall solution at finite radius. This implies 
that the standard value of $\delta_{\rm crit}\approx 1.69$ used in the 
Press-Schechter approximation for the halo mass function at a given time,
based upon extrapolating the linear growth to the epoch at 
which the nonlinear top-hat solution predicts infinite density, should
perhaps be replaced by $(\delta_{\rm L,cross})_{\rm TIS}>1.56$.
A comparison of the TIS model with the SUS and SIS approximations
is summarized in Table~\ref{compare}.
\begin{table}
\begin{minipage}{\textwidth}
\caption{A comparison of three approximations for the postcollapse equilibrium
structure of top-hat density perturbations in a low density universe.}
\label{compare}
\begin{tabular}{@{}lcccc}
&Uniform&Singular& TIS\\
& Sphere\footnote{A top-hat perturbation of a given mass collapses at a given redshift
in a background universe with given values of $\Omega_0$ and $\lambda_0$; all of these 
values are held fixed in this comparison of the three approximations.}
		&Isothermal&Solution\\
&& Sphere\footnote{These SIS numbers are an approximation which ignores the small
modification of the Lane-Emden equation solution to take account of $\Lambda\neq0$, but
accounts for the more important effects of $\Lambda\neq0$ on top-hat evolution,
energy conservation and the virial theorem.}
& $\theta=0-0.5$\footnote{These values show the full range allowed for top-hat perturbations which 
collapse at finite time including the distant future (i.e. $\lambda\rightarrow1$).}
\\\hline
$\eta/\eta_{\rm SUS}$&1&0.833&1.11--1.04\\[3mm]
$\displaystyle{\frac{T}
	{T_{\rm SUS}}}$&1&3&2.16--2.26\\[5mm]
$\displaystyle{\frac{\rho_0}{\rho_t}}$& 1&$\infty$&514--575\\[3mm]
$\displaystyle{\frac{\langle\rho\rangle}{\rho_t}}$&1&3&3.73--3.59\\[3mm]
$\displaystyle{\frac{r_t}{r_0}}$& -- NA --&$\infty$&29.4--31.7\\[3mm]
$\frac{\Delta_c}{\Delta_{\rm c,SUS}}$&1&
      $\displaystyle{\left(\frac65\right)^3}=1.728$&1.36--1.12\\[5mm]
$K/|W|$&0.5&0.75&0.683-0.707
\\\hline
\end{tabular}
\end{minipage}
\end{table}

\subsection{Summary}
We have generalized the TIS model, derived in Paper I for an 
Einstein-de Sitter universe, to the case of a low-density universe, either 
matter-dominated ($\Omega_0<1,\lambda_0=0$)
or flat with cosmological constant ($\Omega_0+\lambda_0=1$). The 
formalism we have presented can also be used to obtain the corresponding
TIS solutions for other background cosmology models with a component of the
energy density which remains homogeneous on scales relevant to the formation of
virialized halos, such as quintessence. The halo density profile we have thus 
derived has a universal, time-invariant shape in the matter-dominated cases,
when expressed in units of the central density $\rho_0$ with radius in
units of the core radius $r_0$. However, in the presence of a cosmological 
constant, whose importance increases with time, the isothermal Lane-Emden 
equation is modified, and, as a result, this dimensionless TIS density 
profile is no longer time-invariant, but, instead, depends 
on the epoch of collapse. This dependence is relatively weak, except for the 
outer parts of the halo at late times. For example, for $\lambda_0=0.7$ and 
$z_{\rm coll}=0$, the dimensionless radius of the halo $r_t/r_0$ is
30.04, larger than the value of 29.40 for a matter-dominated universe
by about 2\% (see Table~\ref{summary_low}). At later times (i.e. in the future)
in the $\Lambda$-dominated universe, however, the departure from the universal, 
time-invariant shape of the matter-dominated case will be more 
pronounced, with the dimensionless radius reaching 31.69, or $\sim8\%$ higher.
For all low-density universe cases, including those which are matter-dominated,
 far more significant differences from the TIS model in an EdS universe, however,
are found in the over-all amplitude of the density profile as measured relative to 
the mean density of the background. For example, the ratios of the
average density inside the virial radius and of the central density to the cosmic mean
background density at the epoch of collapse, are larger then their values for an
EdS universe by more than a factor of 2, for the cases $\Omega_0=0.3,\lambda_0=0$ and
$1-\Omega_0=\lambda_0=0.7$, for halos which collapse today, while these 
ratios grow arbitrarily large 
in the future, due to the decrease of $\Omega(t)$ with time in both cases.

The dimensional parameters of the TIS halo solution also depend significantly on the
background cosmology. For
example, a TIS halo of $10^{10} h^{-1} M_\odot$ which collapses at $z=0$ in an EdS
universe has a radius $r_t=40.35\,h^{-1}{\rm kpc}$, central density 
$\rho_0 = 3.38\times10^{-25}\, h^2 {\rm g\, cm^{-3}}$, and velocity dispersion 
$\sigma_V=22.6\,{\rm km\, s^{-1}}$, while a halo of the same mass which collapses at the 
same redshift, but in a flat universe with $\lambda_0=0.7$, is $\sim 18\%$ larger, with a
$33\%$ lower central density, and $\sim9\%$ lower velocity dispersion.
The same halo in an open universe with $\Omega_0=0.3$ would be $\sim14\%$ larger than in the 
EdS case, with $38\%$ lower central density, and 6\% lower velocity dispersion.

Our results demonstrate that 
the presence of a cosmological constant can influence the 
internal structure of virialized haloes, particularly in their outer regions. 
This is true even if we use a conservative estimate of the contribution of the 
cosmological constant to the mean energy density of the universe, for haloes collapsing 
at the present. In principle, this effect grows in importance for haloes which are 
destined to form in the future.

The TIS model has many characteristics in common with the haloes with 
uniform-density cores which are expected to form from self-interacting dark matter (SIDM)
(Burkert 2000, Dav\'e et al. 2000, Firmani et al. 2000). 
Since our model describes the final equilibrium 
state of a collapsing halo and is based
on a minimum-energy principle, it may transcend the details of the
particle interactions which lead to this equilibrium. If so, then it is
reasonable to expect that our TIS solution will be a useful approximation 
for the equilibrium structure of SIDM haloes. A detailed comparison between
the TIS model haloes derived here and the predictions of halo formation
in the SIDM model would, therefore, be of interest.

The application of the TIS model presented here to the problem of halo formation 
in the CDM model, including further comparisons with N-body simulation results, 
will be described elsewhere. In one such application, the TIS model is shown to 
provide a good theoretical explanation for the observed rotation curves of dark 
matter -- dominated galaxies and the correlation which has been reported between 
the maximum velocity on a given rotation curve and the galactocentric radius at 
which it occurs \cite{ISa}.

\section*{Acknowledgments}

This research was supported in part by NSF grant INT-0003682 from 
the International Research Fellowship Program and the Office of 
Multidisciplinary Activities of the Directorate for Mathematical and
Physical Sciences to ITI
and grants NASA ATP NAG5-7363 and NAG5-7821, NSF ASC-9504046, and Texas 
Advanced Research Program 3658-0624-1999 to PRS.

\appendix
\section{Application of the Self-similar, spherical Infall Solution at early 
times in a low-density universe}
\label{appendix1}

The self-similar cosmological infall solution of Bertschinger (1985) can be
generalized to the case of a low-density universe at early times by 
 simple scalings of the variables. At early times, $\Omega(z)\approx1$, and
both the background universe and density fluctuations evolve approximately
as in an EdS universe. Therefore, the self-similar infall solution 
will be approximately valid. 
Consider the spherical mass shell which is just turning around at some redshift 
$z$. Let the radius of this shell be $r_{\rm ta}$. According to the top-hat
solution applied to this spherical shell, the mass $m_{\rm ta}$ enclosed by 
this shell is just
\begin{equation}
\label{mass_ta}
m_{\rm ta}=\frac43\pi r_{\rm ta}^3{\rho_b(z)}(1+\delta_m),
\end{equation}
where $\rho_b(z)$ is the mean background matter density at redshift $z$ and
$(1+\delta_m)$ is the average overdensity inside this sphere at its epoch of 
maximum expansion. At early times, for which the behaviour approaches that of 
an EdS universe, $(1+\delta_m)=9\pi^2/16$ and 
$\rho_b(z)=\rho_{\rm crit}(z)=3H^2(z)/(8\pi G)$. However, for a low-density
universe at high redshift, the Friedmann equation~(\ref{Friedmann}) 
simplifies to
\begin{equation}
\label{friedmann_approx}
H^2(z)=H_0^2\Omega_0(1+z)^3,
\end{equation}
so
\begin{equation}
\label{rho_crit_low}
\rho_{\rm crit}(\Omega_0,z)=\Omega_0\rho_{\rm crit,EdS}(z).
\end{equation}
Therefore, if we fix the mass inside $r_{\rm ta}$ at redshift $z$, then we 
must take a different value for the radius $r_{\rm ta}$ for the EdS and 
low-density universes, according to equations (\ref{mass_ta}) and 
(\ref{rho_crit_low}), which yields
\begin{equation}
\label{r_ta_ratio}
r_{\rm ta}(z)/r_{\rm ta,EdS}(z)
	=[\rho_{\rm crit}(\Omega_0,z)/\rho_{\rm crit,EdS}(z)]^{-1/3}=\Omega_0^{-1/3}.
\end{equation}
Consider now the mass within a sphere whose radius $r$ at this redshift $z$ is
a fixed fraction $\lambda_B$ [in the notation of Bertschinger (1985), where 
we have added the subscript ``B'' to this dimensionless radius coordinate 
to  distinguish it here from the unrelated quantities 
$\lambda_0$ and $\lambda(t)$ used elsewhere in this paper to refer
 to the cosmological constant in units of the critical density,
as measured at the present and at time t, respectively] of the 
radius $r_{\rm ta}$,
\begin{equation}
\label{mass_bert}
m(\lambda_B,z)=m_{\rm ta}(z)\frac{M(\lambda_B)}{1+\delta_m},
\end{equation}
where $M(\lambda_B)$ is a dimensionless function of $\lambda_B$ only, for 
which $M(1)=1+\delta_m$.
According to equation~(\ref{mass_bert}), if we fix the mass $m_{\rm ta}$
inside $r_{\rm ta}$ at redshift $z$
when comparing the EdS and low-density universe solutions, we must also have 
the same mass $m(\lambda_B,z)$ inside any $\lambda_B$ in the two solutions.
In that case, the radius $r$ for any $\lambda_B$ must correspondingly 
relate to the radius $r_{\rm EdS}$ which encloses the same mass at the
same redshift, according to 
\begin{equation}    
\label{bert_rad}
r(\lambda_B)=r_{\rm EdS}(\lambda_B)\Omega_0^{-1/3}.
\end{equation}
The age of a low-density universe at a given redshift is related
to the age of an EdS universe at the same redshift according to the 
solutions of the Friedmann equation for these two 
cases. Equation~(\ref{friedmann_approx}) is easily integrated analytically
to yield the result for high redshift, 
\begin{equation}
\label{bert_time}
t(\Omega_0,z)=t_{\rm EdS}(z)\Omega_0^{-1/2}.
\end{equation}
This determines the scaling of velocities in the infall solution, as
follows. At any $\lambda_B$, the velocity scales as
\begin{equation}
\label{vel_scaling}
v\propto\frac{r}{t}  
\end{equation}
Equations~(\ref{bert_rad}) -- (\ref{vel_scaling}) yield
\begin{equation}
\label{bert_vel}
v(\lambda_B)=v_{\rm EdS}(\lambda_B)\Omega_0^{1/6}. 
\end{equation}

\section{Analytical fitting formulae for the dependence 
of the TIS halo dimensionless parameters on $\theta$ and the
dependence of the density profile on the dimensionless radius $\zeta$}
\label{appendix2}

Analytical fitting formulae which closely approximate the numerical results
for the dimensionless TIS parameters will make applications of our TIS model much
more convenient. We have derived such fitting formulae for two overlapping 
intervals of $\theta$, with different fractional errors in these two intervals, 
as follows.

In the currently most relevant cosmological range, $0\leq\theta\leq 0.123$, the fits are
given by
\begin{eqnarray}
\label{fits3_zeta_t}
\zeta_t   \!\!\! \!       &=& \!\! \!\!29.4003+11.4652\theta-13.8428\theta^2-12.8453\theta^3,\\
\alpha    \!\!\! \!       &=& \!\! \!\!3.7296-0.866069\theta+1.92742\theta^2-1.70326\theta^3,\\
\eta_{\rm TIS}\!\!\! \!   &=&  \!\!\!\! 0.554384-0.45529\theta+0.21258\theta^2+0.02128\theta^3,\\
{\cal R}  \! \!\!\!&=& \! \!\!\!513.842+282.031\theta-193.617\theta^2-748.254\theta^3,\\
\tilde{\rho}_{\lambda}\!\! \!\!\!&=&
		\!\!\!0.00123263\theta-0.00385345\theta^2+0.00579887\theta^3,\\
\label{fits3_Mtil_t}
\tilde{M}_t     \!\!\! \! &=& \! \!\!\!61.485+23.8887\theta-32.9854\theta^2-14.0272\theta^3.
\end{eqnarray}
The relative errors of these fits are $<0.01\%$ (except for the fit to $\tilde{\rho}_{\lambda}$, 
for which the relative error as $\theta\rightarrow0$ is higher, since the exact solution has a 
slightly different slope at $\theta=0$ from that of the fitting formula,  
$\tilde{\rho}_{\lambda}\rightarrow a\theta$, where $a\approx0.001217$. 
However,  for $\theta<0.005$ where the above fit for $\tilde{\rho}_\lambda$
is not perfect, the departure from the result for the EdS case is negligible anyway.)
   
The fits in the full allowed range $0\leq\theta\leq 0.5$ are given by
\begin{eqnarray}
\label{fits4_zeta_t}
\zeta_t   \!\!\! \!       &=& \!\! \!\!29.3893+12.0474\theta-20.752\theta^2+11.732\theta^3,\\
\alpha    \!\!\! \!       &=& \!\! \!\! 3.72913-0.843068\theta+1.71017\theta^2-1.20963\theta^3,\\
\eta_{\rm TIS}        \!\!\! \!     &=&  \!\!\!\!0.554437 -0.45804\theta+0.23999\theta^2-0.04279\theta^3,\\
{\cal R}  \! \!\!\!&=& \! \!\!\!513.366+306.812\theta-478.871\theta^2+222.726\theta^3,\\
\tilde{\rho}_{\lambda}\!\! \!\!\!&=&
		\!\!\!10^{-3}(
	1.1156\theta-2.6093\theta^2+2.1242\theta^3),\\
\label{fits4_Mtil_t}
\tilde{M}_t     \!\!\! \! &=& \! \!\!\! 61.4683+24.782\theta-43.8578\theta^2+25.8869\theta^3.
\end{eqnarray}
The relative errors of these fits are $<0.1\%$ (again with the exception of the 
fit for $\tilde{\rho}_{\lambda}$).

In Paper I we obtained a simple analytical fit to the TIS density
profile, as follows: 
\begin{equation}
\label{rho-analyt_low}
\tilde{\rho}(\zeta)=\displaystyle{\frac{A}{a^2+\zeta^2}
    -\frac{B}{b^2+\zeta^2}},
\end{equation}
for $\zeta\leq\zeta_t$, with
\begin{equation}
\label{analyt_our_low}
(A,a^2,B,b^2)_{TIS}=(21.38,9.08,19.81,14.62).
\end{equation}
This fit is accurate to within 3\% inside the virial radius for $\lambda_0=0$.
In the presence of a cosmological constant, the dimensionless profile varies, as 
described above. Nevertheless, for $\lambda_0<0.9$ and haloes that collapse by 
$z=0$, the fit still has accuracy of better than 3\%. Overall the fit deteriorates 
a bit, however, by slightly overestimating $\tilde{\rho}$ at all radii.

The corresponding fit to the TIS halo circular velocity profile (i.e. the 
rotation curve), obtained simply by integrating equation~(\ref{rho-analyt_low}), is 
\begin{eqnarray}
\label{v_analyt}
\displaystyle{v_c(r)
=\sigma_V\left\{A-B
	+\frac{r_0}{r}\left[bB\arctan\left(\frac{r}{br_0}\right)
	-aA\arctan\left(\frac{r}{ar_0}\right)\right]\right\}^{1/2}}
\end{eqnarray}
\cite{ISa}.
This fit to $v_c(r)/\sigma_V$ is good to $<1\%$ for $\lambda_0=0$, and for $r<2r_t/3$
if $\lambda_0\neq0$, while at larger radii, depending on $\lambda_0$ and $z_{\rm coll}$, the
error could be larger but is not greater than 6\% at $r_t$, even for $\lambda_0=0.9$ and 
$z_{\rm coll}=0$. Most of the departure of the exact circular velocity profile
from this fit is due
to the $\rho_\lambda$-dependent correction factor in $v_c$, which was omitted
here for simplicity. In Iliev \& Shapiro (2001), we apply the TIS model to
explain the observed rotation curves of dark matter--dominated galaxies and the statistical 
correlations amongst their rotation curve parameters for different mass haloes collapsing
at different epochs in the CDM model for different background cosmologies.
\end{document}